\documentclass[%
superscriptaddress,
showpacs,
nofootinbib,
amsmath,
amssymb,
aps,
pra
]{revtex4-1}

\usepackage{graphicx}
\usepackage{dcolumn}
\usepackage{bm}
\usepackage{hyperref}

\usepackage{amsthm}
\usepackage{epstopdf}
\usepackage{subfigure}
\usepackage{enumerate}
\usepackage{ragged2e}

\usepackage{newfloat}
\usepackage{algcompatible}
\DeclareFloatingEnvironment[
listname=List of Algorithms,
name=ALG.,
placement=tbhp]{algorithm}

\usepackage{mdframed}
\newmdenv[leftline=false,rightline=false,innertopmargin=2pt,innerbottommargin=2pt]{topbot}

\DeclareFloatingEnvironment[
	listname=List of Algorithms,
	name=PROG.,
	placement=tbhp]{program}
\usepackage{mmacells}
\usepackage[abs]{overpic}

\newcommand{\aref}[1]{\hyperref[#1]{Appendix~\ref{#1}}}

\theoremstyle{plain}

\theoremstyle{definition}

\theoremstyle{remark}

\newcommand{\ket}[1]{\ensuremath{\left|#1\right\rangle}}

\begin{document}
	
	\preprint{APS/123-QED}
	
	\title{Zero Transfer in Continuous Time Quantum Walks}
	
	\author{A. Sett}
\affiliation{Department of Physics, The University of Western Australia, Perth, WA 6009, Australia}
    \author{H. Pan}
\affiliation{Department of Physics, The University of Western Australia, Perth, WA 6009, Australia}
    \affiliation{Department of Physics, University of Maryland, College Park, MD 20742, USA}
	\author{P. E. Falloon}
\affiliation{Department of Physics, The University of Western Australia, Perth, WA 6009, Australia}
	\author{J. B. Wang}
	\email{jingbo.wang@uwa.edu.au}
	\affiliation{Department of Physics, The University of Western Australia, Perth, WA 6009, Australia}
	\date{\today}

\title{Zero Transfer in Continuous Time Quantum Walks}

\begin{abstract}
In this paper we show how using complex valued edge weights in a graph can completely suppress the flow of probability amplitude in a continuous time quantum walk to specific vertices of the graph when the edge weights, graph topology and initial state of the quantum walk satisfy certain conditions. The conditions presented in this paper are derived from the so-called chiral quantum walk, a variant of the continuous time quantum walk which incorporates directional bias with respect to site transfer probabilities between vertices of a graph by using complex edge weights.  We examine the necessity to break the time reversal symmetry in order to achieve zero transfer in continuous time quantum walks. We also consider the effect of decoherence on zero transfer and suggest that this phenomena may be used to detect decoherence in the system.
\end{abstract}

\maketitle 
    
\section{Introduction}
Quantum computing and quantum information hold the promise of alternative computing models which, using quantum algorithms, are able to solve problems presently considered to be intractable on classical computers \cite{nielsen2010quantum}. A notable example is Shor's factoring algorithm, which is able to solve the factoring and discrete logarithm problem in polynomial time \cite{shor1994algorithms}. In addition to this, the field has found uses in areas such as quantum key distribution and quantum teleportation \cite{bennett2014quantum,bennett1993teleporting} and it has also been shown that with a universal quantum computer we will be able to simulate local quantum systems efficiently \cite{lloyd1996universal} --- a task seen as computationally infeasible on a classical computer  \cite{feynman1982simulating}. One such model of quantum computing that has been shown to be universal \cite{childs2009universal} is the continuous-time quantum walk (CTQW) model formulated by Farhi and Gutmann \cite{farhi1998quantum}. \\
	\\
	The CTQW is the quantum-mechanical analogue of the classical random walk, in which a ``walker" traverses the vertices of some graph via a sequence of randomly chosen ``steps'' along its edges. The classical random walk forms the basis of many classical algorithms and has been used in a variety of fields such as modeling Brownian motion in physics and algorithms for image colourisation in computer science \cite{einstein1956investigations,liu2009colorization}. In the CTQW, the walker is quantum in nature, in the sense that the walker can be present at several vertices at once with associated probabilities, analogous to how a quantum particle can exist in a superposition of quantum states. Quantum interference between these superposition states results in markedly different probability distributions for the walker's position over time \cite{ambainis2003quantum,wang2013physical}. This unusual property has allowed CTQWs to be extensively utilised in graph theoretical applications \cite{izaac2017centrality,loke2017comparing,gamble2010two,xu2018quantum} and as the basis of several  quantum algorithms \cite{xu2018quantum,childs2004spatial,li2015analytical,qiang2016efficient,twoqubitQiang, childs2003exponential}.\\
	\\
	One limitation of the CTQW is that the transformation describing the evolution of the walker must be unitary. As a result, the CTQW is restricted to undirected graphs and no directional biasing in terms of site transfer probabilities between vertices of the graph can be obtained. To generalise the CTQW model Whitfield et al. formulated the quantum stochastic walk, in which the walker evolves according to the Lindblad equation \cite{whitfield2010quantum}. This model was subsequently applied to directional graphs by letting the unitary part of the equation describe the standard CTQW and using the classical non-unitary part of the equation to incorporate directionality \cite{sanchez2012quantum}. \\
	\\
	More recently, Zimbor\'{a}s et al. have shown that by breaking a time-reversal symmetry (TRS) condition associated with site transfer probabilities in continuous time quantum walks, transport enhancement/suppression in coherent transport network models can be achieved \cite{zimboras2013quantum}. This is done by using complex phase terms as edge weightings for the respective graph such that the site transfer probability between two distinct vertices of the graph depends on which direction the walker traverses. This is known as the ``chiral" quantum walk. In addition to this Lu et al. showed that by breaking an equivalent TRS condition associated with quantum circuits, perfect state transfer can occur \cite{lu2016chiral}. Such a result suggests the possibility of other subroutines for dynamic quantum control \cite{lu2016chiral}.\\
	\\
	In this paper we observe properties of the state vector obtained from the chiral quantum walk and deduce conditions that must be satisfied to completely suppress information transfer to specific vertices of a graph given that its initial state and the graph topology satisfy certain conditions. That is, the probability of observing the walker at the vertex will be zero at all times. We then consider a quantum stochastic walk, and find that the zero transfer phenomenon is no longer observed in the presence of decoherence (dissipation and dephasing scattering). Finally we utilize this sensitivity to decoherence to construct a simple protocol for estimating the magnitude of decoherence in a quantum system.

\section{Theory}
	
\subsection{Graph Theory}
\begin{figure}[h] \centering
    	\subfigure[]{
		\includegraphics[width=0.30\textwidth]{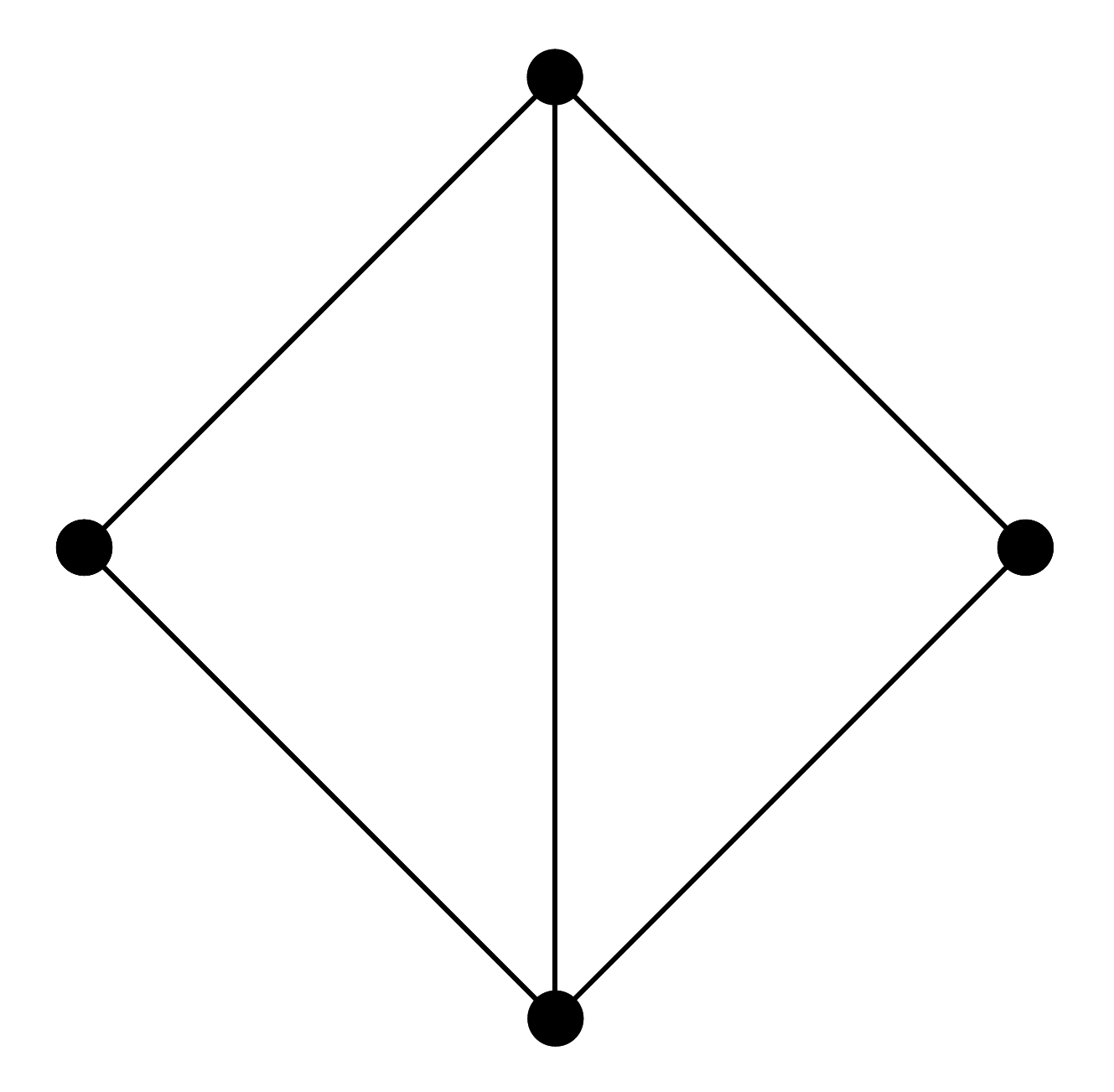}}
         \hspace*{3cm} 
        \subfigure[]{
		\includegraphics[width=0.30\textwidth]{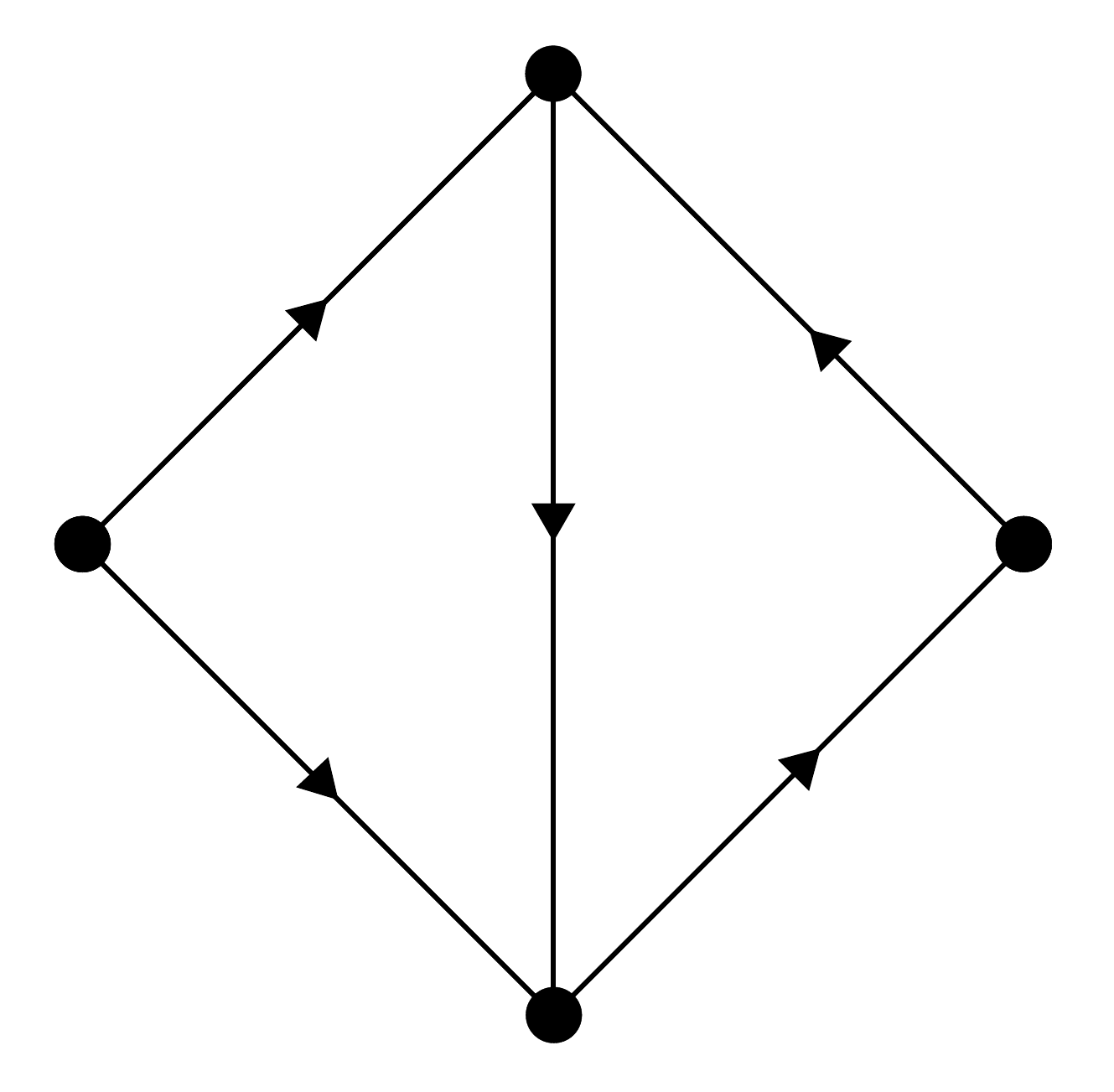}}
		\caption{Examples of an undirected  and directed graph shown in (a) and (b) respectively. Circles, lines and lines with arrows represent vertices, edges and directed edges respectively.} \label{fig:dir-undir}
\end{figure}

\begin{figure}[h] \centering
    	\subfigure[]{
		\begin{overpic}[width=0.30\textwidth]{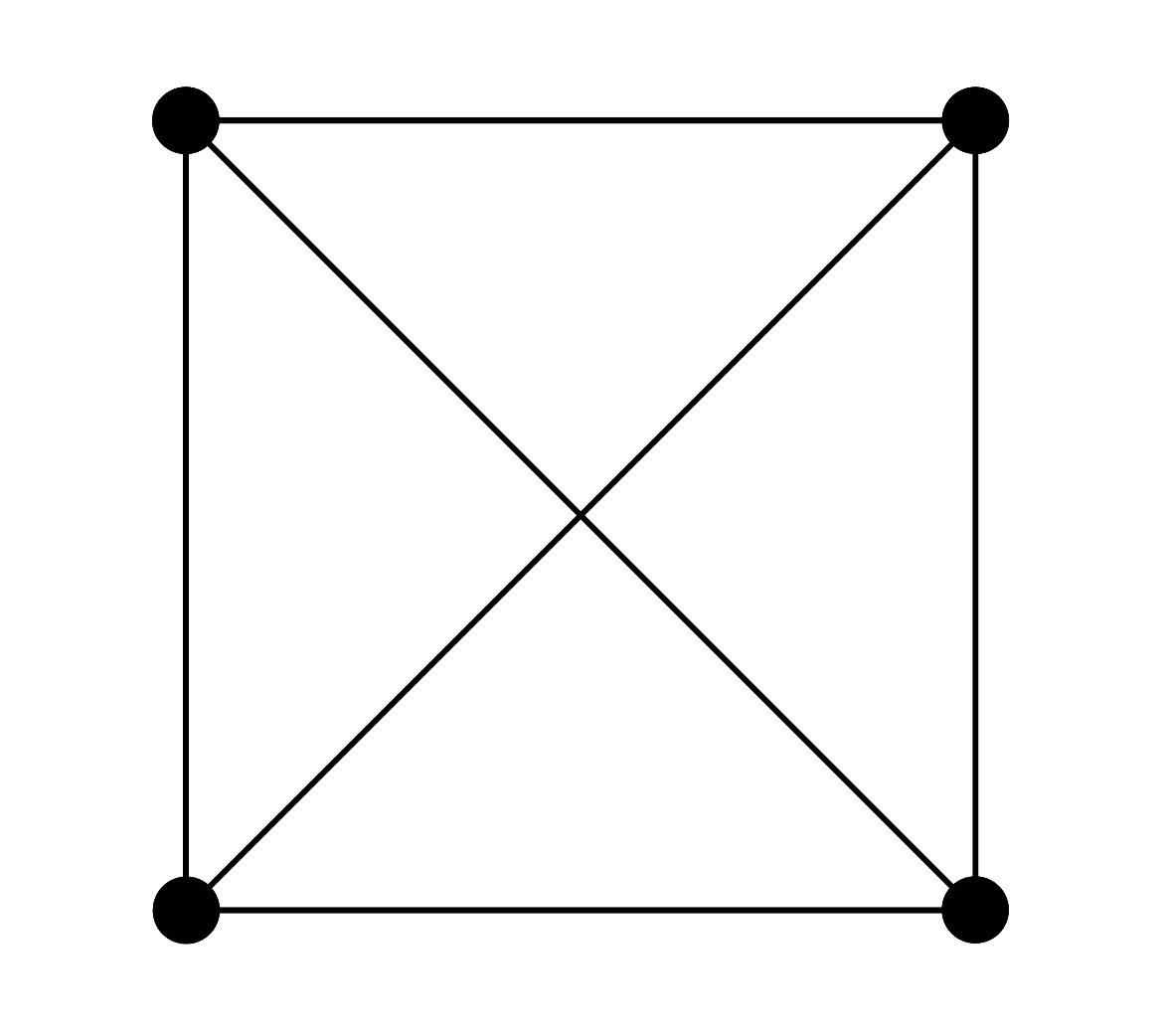}	
		\end{overpic}}
        \hspace*{3cm} 
        \subfigure[]{
		\begin{overpic}[width=0.30\textwidth]{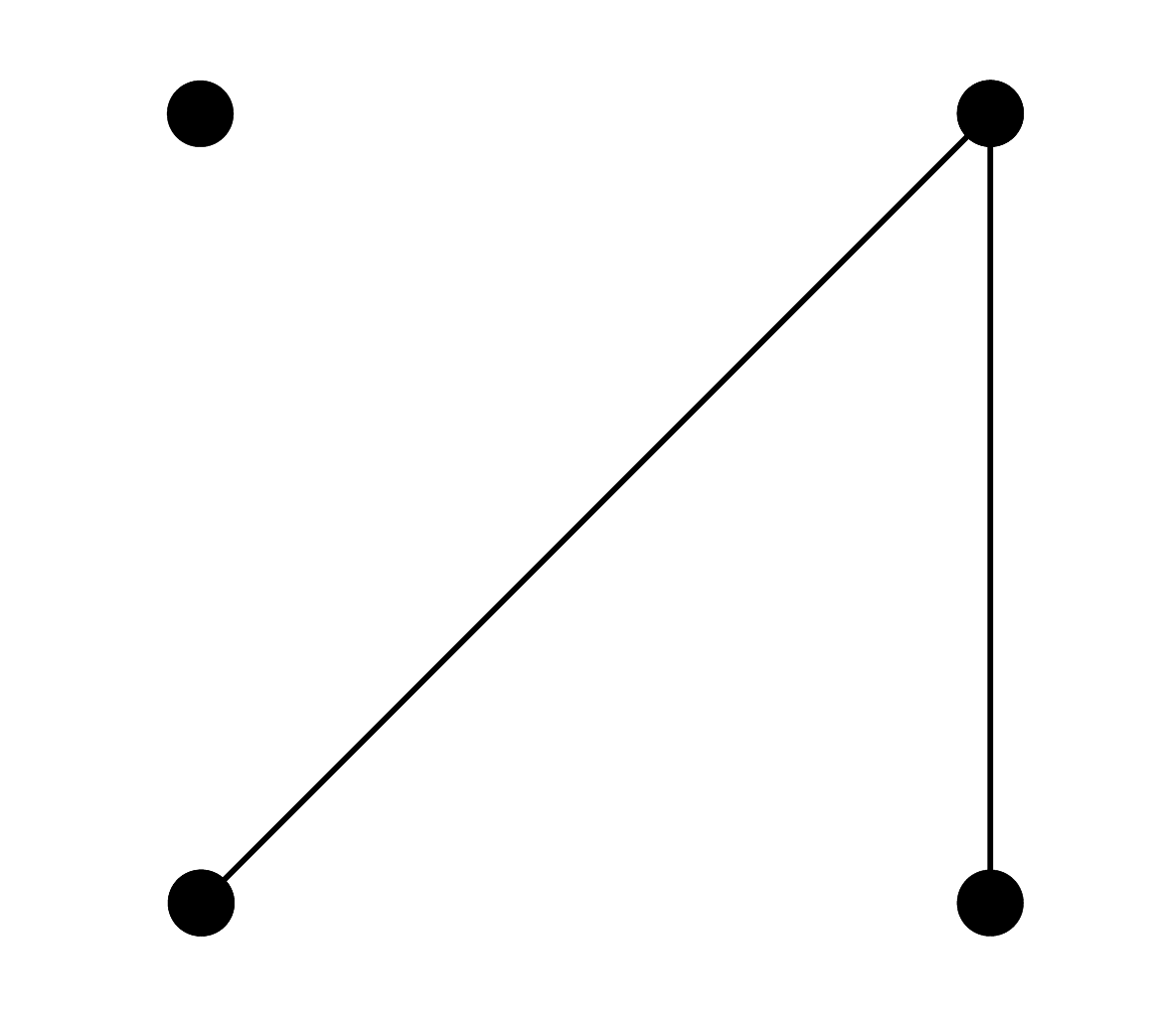}	
		\end{overpic}}
		\caption{Example of a graph (a) and a spanning subgraph (b).}
        \label{fig:span}
	\end{figure}

	We first introduce some elementary graph theory which will be helpful later in describing some of our results. 
Formally, a graph $G$ comprises a vertex set $V$, together with an edge set $E$ containing pairs of vertices $(x,y)$ with $x,y \in V$. For an \emph{undirected} graph, $(x,y)$ is interpreted as an unordered pair, while for a \emph{directed} graph it represents the edge from $x$ to $y$. If an edge exists between the vertices $x$ and $y$, we say that the vertices are adjacent, denoted $x \sim y$ \cite{godsil2013algebraic}. For both undirected and directed graphs the total number of vertices is denoted by $|G|$. Examples of an undirected and directed graph are shown in figure \ref{fig:dir-undir}. For the remainder of this section we will look at the properties of undirected graphs; however, all properties mentioned apply to directed graphs as well. \\
\\	
	\\
	A \emph{subgraph} of a graph $G$ is a graph $\Gamma$ satisfying $V(\Gamma) \subseteq V(G)$ and $E(\Gamma) \subseteq E(G)$. A \emph{spanning} subgraph is one for which $V(\Gamma) = V(G)$. A spanning subgraph can be obtained from a graph by deleting only edges of the graph \cite{godsil2013algebraic}. Examples of a graph and a spanning subgraph are shown in figure \ref{fig:span}.
\\
	\\
	\\ The adjacency matrix, $A$, of a graph with $|G|=n$ vertices is an $n \times n$ matrix where the $ij^{th}$ entry, $A_{ij}$, represents the edge $(i,j)$. If $(i,j) \in E(G)$ then $A_{ij}=1$. If there are multiple edges between vertices, then $A_{ij}=m$, where $m$ is the number of edges between $i$ and $j$. Edges can also have weights which can be real or complex valued. If $(i,j) \in E(G)$ and has a weighting of $w$ then $A_{ij}=w$ \cite{von2007tutorial}.  Note for  graphs, which are unweighted, undirected and have no multiple edges, the adjacency matrix will be real-symmetric and only consist of $1$'s and $0$'s. \\
	\\
	A \emph{complete} graph is an unweighted, undirected graph where every vertex is adjacent to every other vertex of the graph (excluding itself). The complete graph consisting of $n$ vertices is denoted $K_n$. A \emph{path} is a sequence $x\sim y\sim ...\sim z$ of distinct vertices, with each vertex adjacent to the previous one. A \emph{cycle} is a path where the first and last vertex in the sequence that describes the path is the same. The cycle graph with $n$ vertices is denoted $C_n$ \cite{godsil2013algebraic}.
	
	\subsection{Chiral Quantum Walks}
\begin{figure}[h] \centering
    	\subfigure[]{
		\begin{overpic}[width=0.35\textwidth]{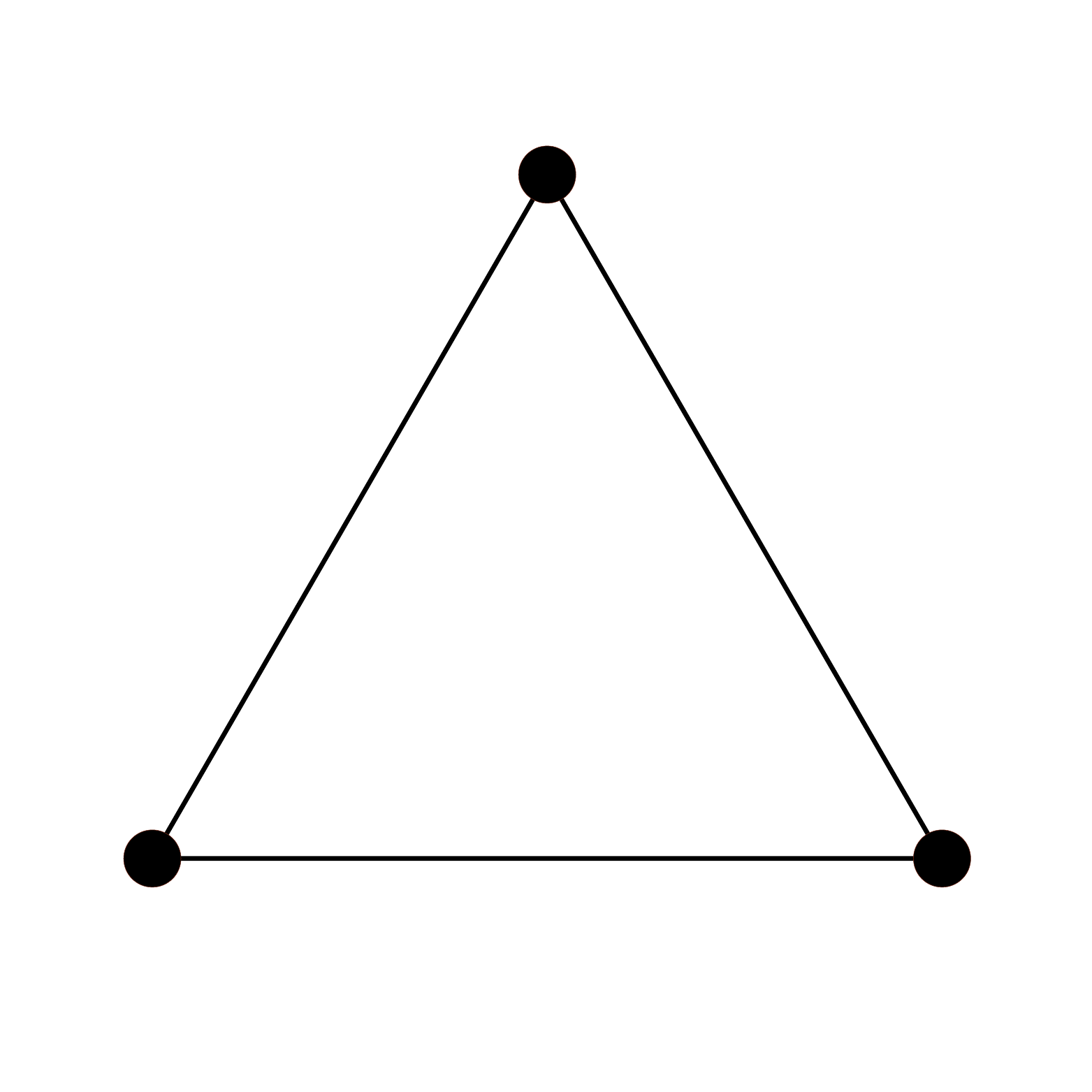}	
        \put(88,160){1}
        \put(13,37){2}
        \put(163,37){3}
		\end{overpic}}
        \hspace*{3cm} 
        \subfigure[]{
		\begin{overpic}[width=0.35\textwidth]{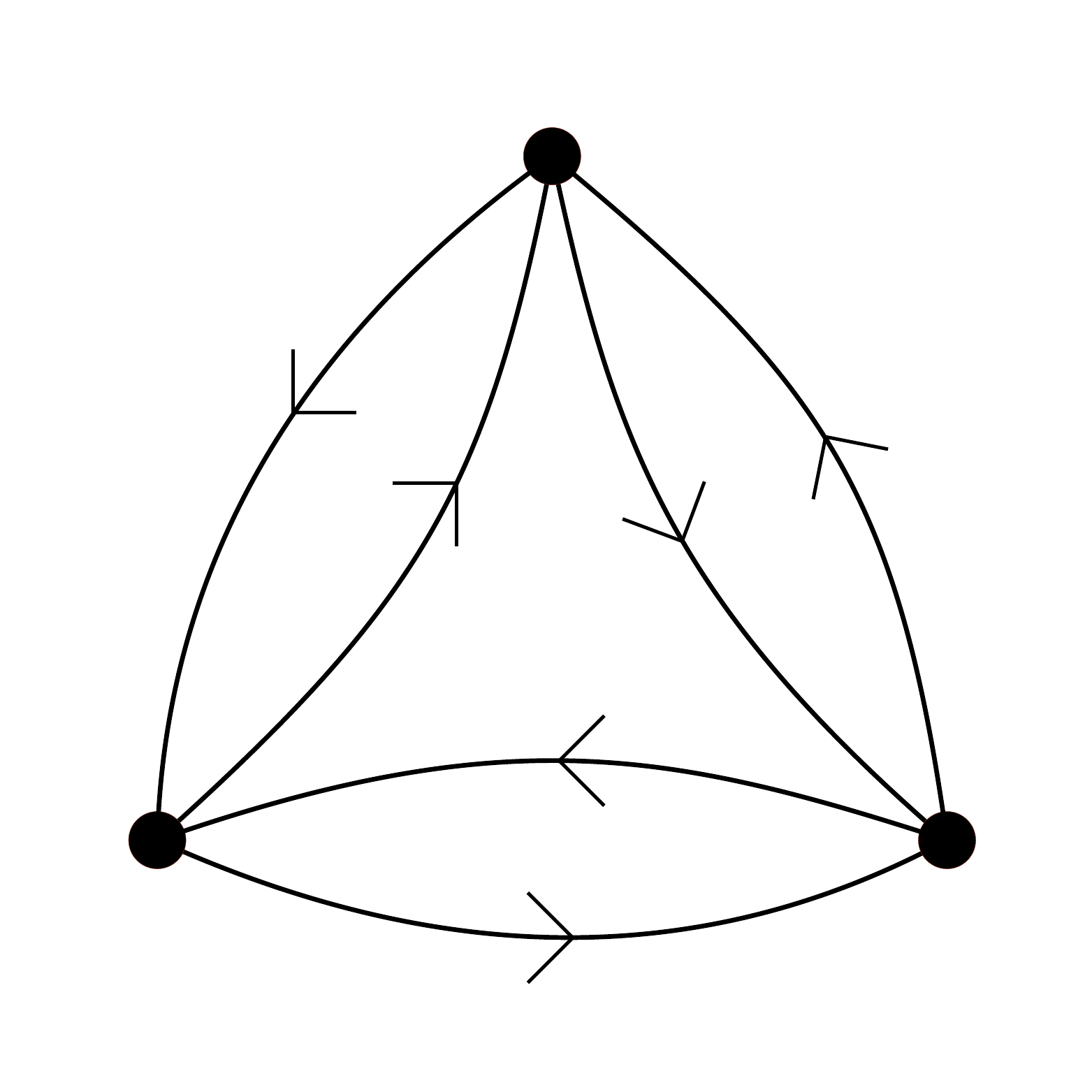}
            \put(89,162){1}
        	\put(14,37){2}
        	\put(164,37){3}
        	\put(36, 120){$e^{i\alpha_1}$}
			\put(46,85){$e^{-i\alpha_1}$}
			\put(150, 80){$e^{i\alpha_2}$}
			\put(110,105){$e^{-i\alpha_2}$}
			\put(90, 10){$e^{i\alpha_3}$}
			\put(90,60){$e^{-i\alpha_3}$}
		\end{overpic}}
		\caption{$K_3$ and the corresponding directed graph with complex edge weights shown in figures (a) and (b) respectively.} \label{fig:chiK3}
	\end{figure}
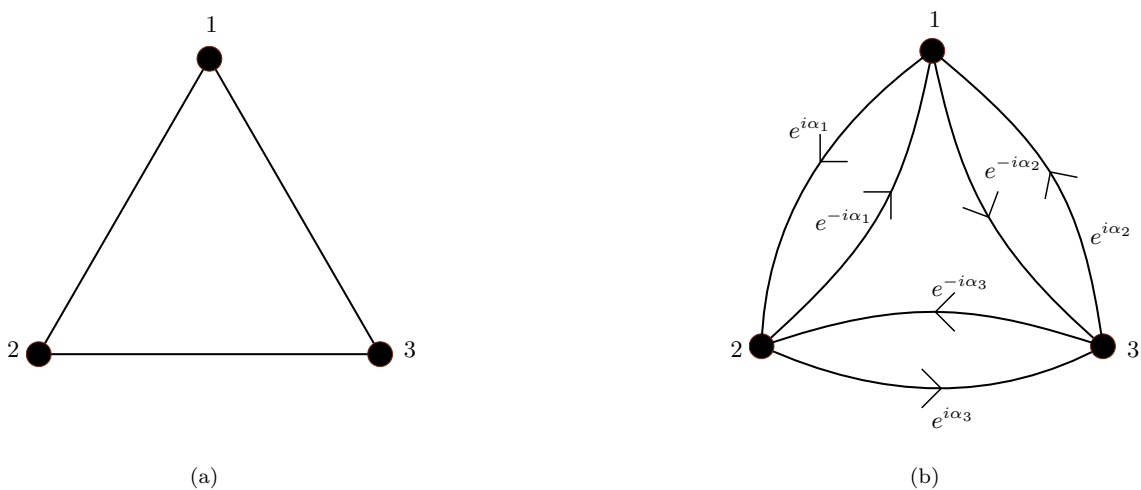

	The continuous-time quantum walk is the quantum analogue of the continuous time classical random walk \cite{ambainis2003quantum}. Given a graph consisting of $|G|=n$ vertices, the corresponding state vector, $| \psi(t) \rangle$, is represented in the $n$-dimensional orthonormal basis $\{|1\rangle,|2\rangle,...,|n\rangle\}$, where $|i\rangle$ denotes a state occupying vertex $i$. $| \psi(t) \rangle$ evolves from an initial state, $|\psi(0)\rangle$, via the equation $|\psi(t) \rangle = U| \psi(0) \rangle$ where $U = e^{-iHt}$.	Here, $U$ is the time evolution operator and $H$ is the time-independent Hamiltonian of the system. The postulates of quantum mechanics dictate that the Hamiltonian, $H$, of a quantum system must be Hermitian, implying that $U$ will be unitary \cite{nielsen2010quantum}. For CTQWs, $H$ is usually defined in terms of the adjacency matrix of $G$ \cite{ambainis2003quantum}. \\
	\\
	We now consider the case where time, $t$, flows in the opposite direction. The time-reversed CTQW is given by the equation $|\psi(-t)\rangle = e^{iHt}|\psi(0)\rangle$ and therefore $ |\psi(-t)\rangle  =U^\dag|\psi(0)\rangle$. Next, consider the site transfer probability (STP) between arbitrary vertices of the graph $i$ and $j$. That is, if the walker is at vertex $i$ we wish to find the probability it will transfer to vertex $j$. The STP from vertex $i$ to vertex $j$ is given by $|U_{ji}|^2$ whilst the STP from $i$ to $j$ in the time-reversed case is given by $|U_{ij}|^2$. We say that a CTQW exhibits time-reversal symmetry (TRS) when the STPs in the forward and reverse time case between two arbitrary vertices of the graph, $i$ and $j$, are equal \cite{lu2016chiral}. This gives the following TRS condition: $|U_{ji}|^2 = |U_{ij}|^2$. Note that if a CTQW satisfies the TRS condition, since the STP from $i$ to $j$ in the reversed time case is the same as the transfer probability from $j $ to $i$ in the forward time case, this implies that there is no directional bias in regards to STP along any edges of the graph.\\
	\\
Considering a CTQW on an unweighted, undirected graph, we have $ U =e^{-iAt} = \sum_{k=0}^\infty \frac{(-iAt)^k}{k!}$. It can be shown that all terms of this sum will be symmetric and hence $U$ must be symmetric too. Hence, we have $|U_{ji}|^2 = |U_{ij}|^2$.
Therefore, the TRS condition holds for all unweighted, undirected graphs. This poses an immediate problem when trying to incorporate directional bias in a quantum walk when we choose the Hamiltonian of the walk to be the adjacency matrix of the graph.  It is also noted that if the Hamiltonian is  chosen to be the adjacency matrix, then one can only consider CTQWs on undirected graphs since otherwise its hermicity requirement will be violated. 
In addition to incorporating directionality in CTQWs previous efforts have been made to also establish dynamic control over the directional biasing of STP in a given graph \cite{lu2016chiral}. \\
\\
To incorporate directional bias in STPs and break the TRS condition in a CTQW, Zimbor\'{a}s et al. propose multiplying graph edge weightings with complex phase terms, $e^{i\alpha}$. In order to satisfy the hermicity condition of $H$, the complex conjugate of this term must be multiplied on the diagonally opposite entry of the corresponding adjacency matrix resulting in a directional graph with complex edge weights. An example of this phase appending to $K_3$ is shown in figure \ref{fig:chiK3}. This appendage of complex edge weights to break TRS is known as the ``chiral" quantum walk (CQW) \cite{zimboras2013quantum}.\\
    \\
Previously the CQW has been shown to be able to achieve transport enhancement and suppression in various graphs \cite{zimboras2013quantum}. It has also been shown that, for certain graph topologies, the TRS condition will unconditionally hold despite the addition of phased appended graph edge weightings, and that perfect state transfer can be implemented in quantum circuits using TRS breaking methods \cite{lu2016chiral}. 
  

	\section{Zero Transfer Conditions}
\begin{figure}[h] \centering
		\begin{overpic}[width=0.6\textwidth]{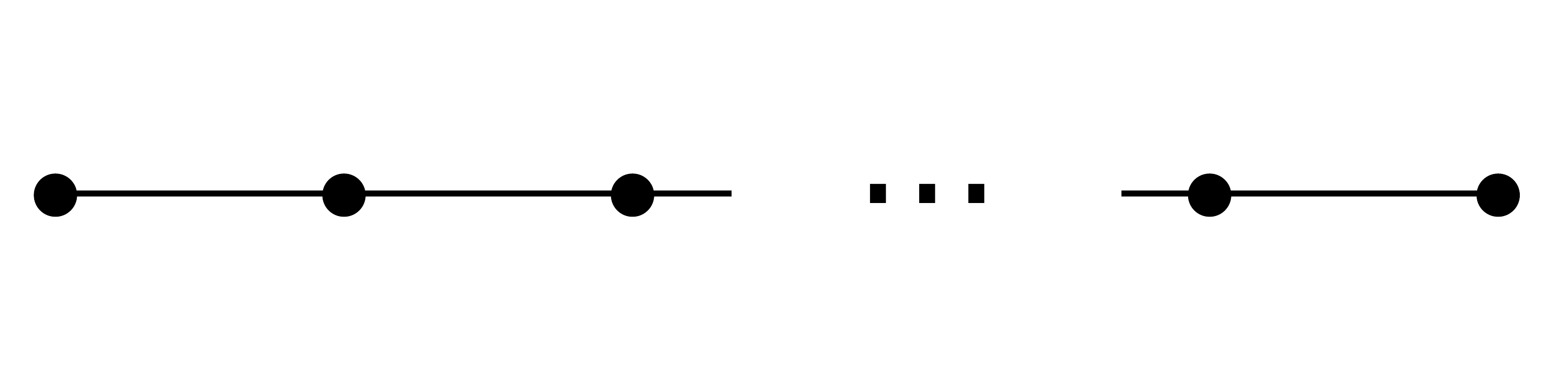}
			\put(8, 55){$1$}
			\put(65, 55){$2$}
			\put(122, 55){$3$}
			\put(226, 55){$n-1$}
			\put(292, 55){$n$}
		\end{overpic}
		\caption{Path with $n$ vertices.} \label{fig:path}
	\end{figure}
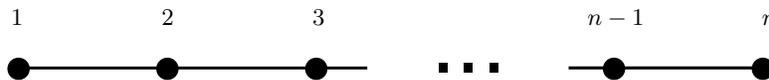

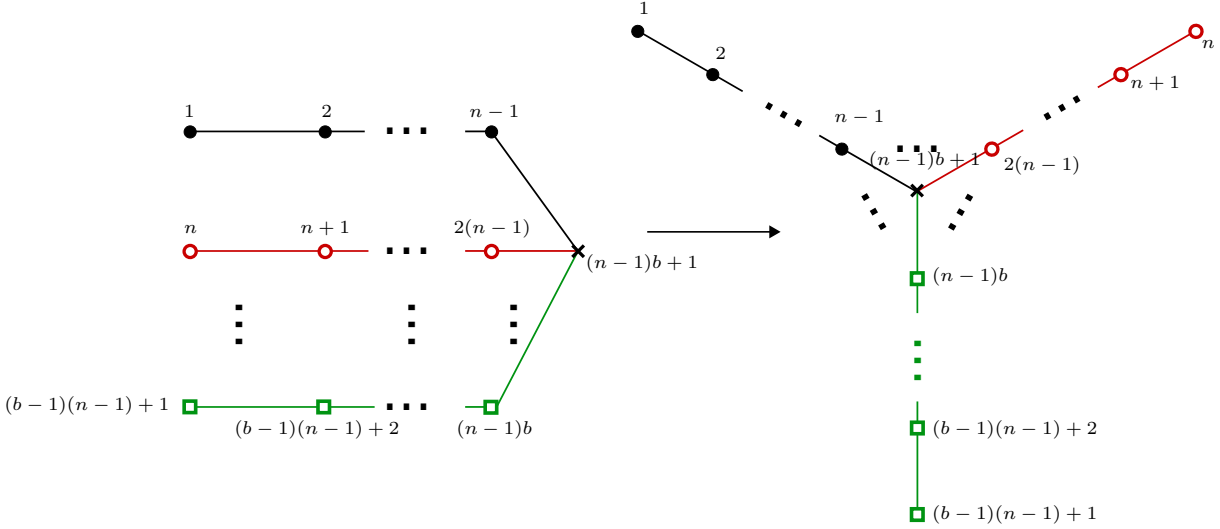
\begin{figure}[htb]
		\centering
		\begin{overpic}[width=0.77\textwidth]{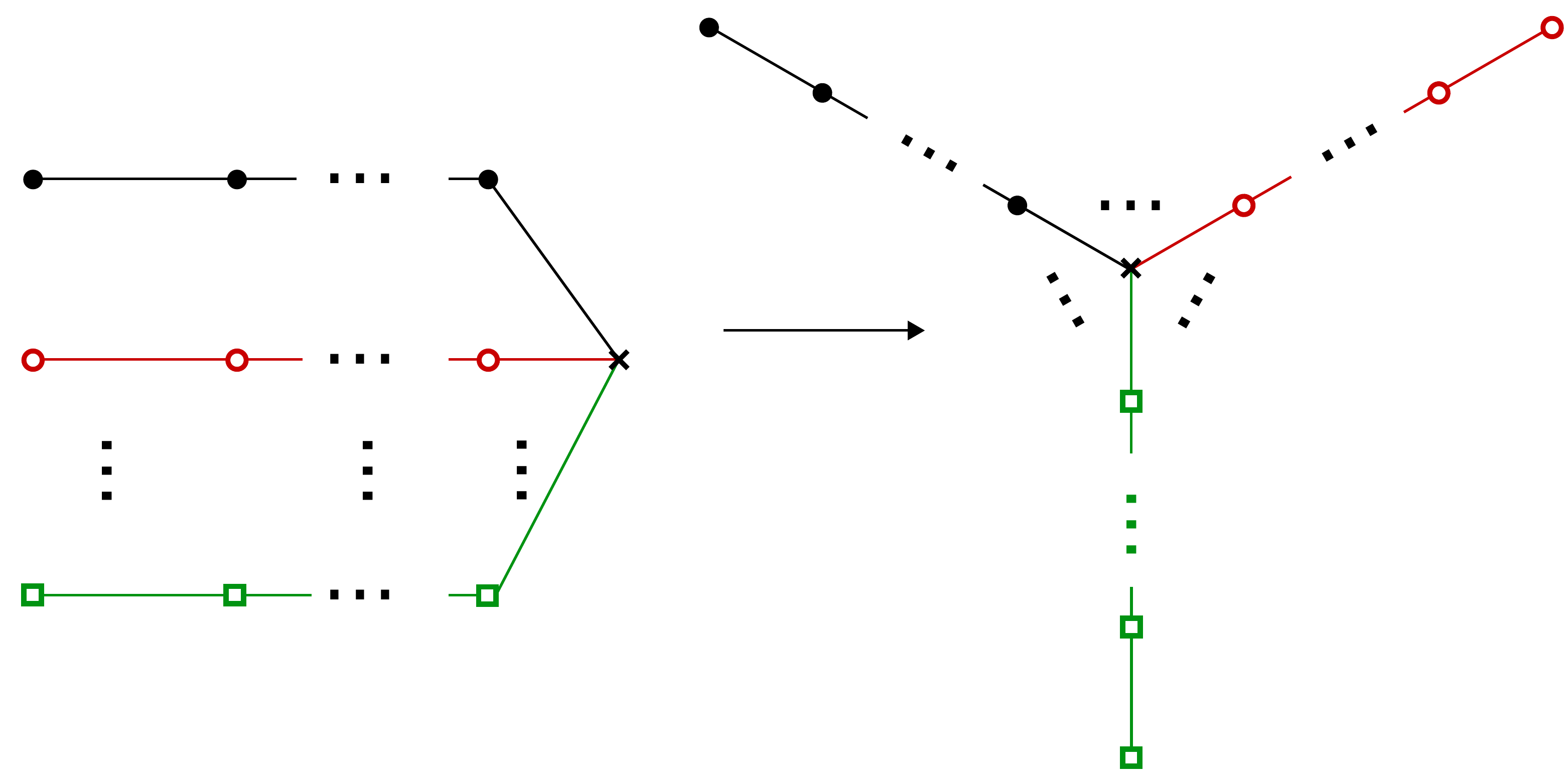}
			\put(6,157){\fontsize{7}{1}$1$}
			\put(58,157){\fontsize{7}{1}$2$}
			\put(114,157){\fontsize{7}{1}$n-1$}
			\put(6,113){\fontsize{7}{1}$n$}
			\put(50,113){\fontsize{7}{1}$n+1$}
			\put(108,113){\fontsize{7}{1}$2(n-1)$}
			\put(158,100){\fontsize{7}{1}$(n-1)b+1$}
			\put(-62,46){\fontsize{7}{1}$(b-1)(n-1)+1$}
			\put(25,37){\fontsize{7}{1}$(b-1)(n-1)+2$}
			\put(109,37){\fontsize{7}{1}$(n-1)b$}
			\put(178,196){\fontsize{7}{1}$1$}
			\put(207,179){\fontsize{7}{1}$2$}
			\put(252,155){\fontsize{7}{1}$n-1$}
			\put(391,183){\fontsize{7}{1}$n$}
			\put(364,168){\fontsize{7}{1}$n+1$}
			\put(316,137){\fontsize{7}{1}$2(n-1)$}
			\put(289,5){\fontsize{7}{1}$(b-1)(n-1)+1$}
			\put(289,37){\fontsize{7}{1}$(b-1)(n-1)+2$}
			\put(289,95){\fontsize{7}{1}$(n-1)b$}
			\put(265,139){\fontsize{7}{1}$(n-1)b+1$}
		\end{overpic}
		\caption{Graph formed from merging $b$ paths where every path consists of $n$ vertices.} \label{fig:type1}
	\end{figure}

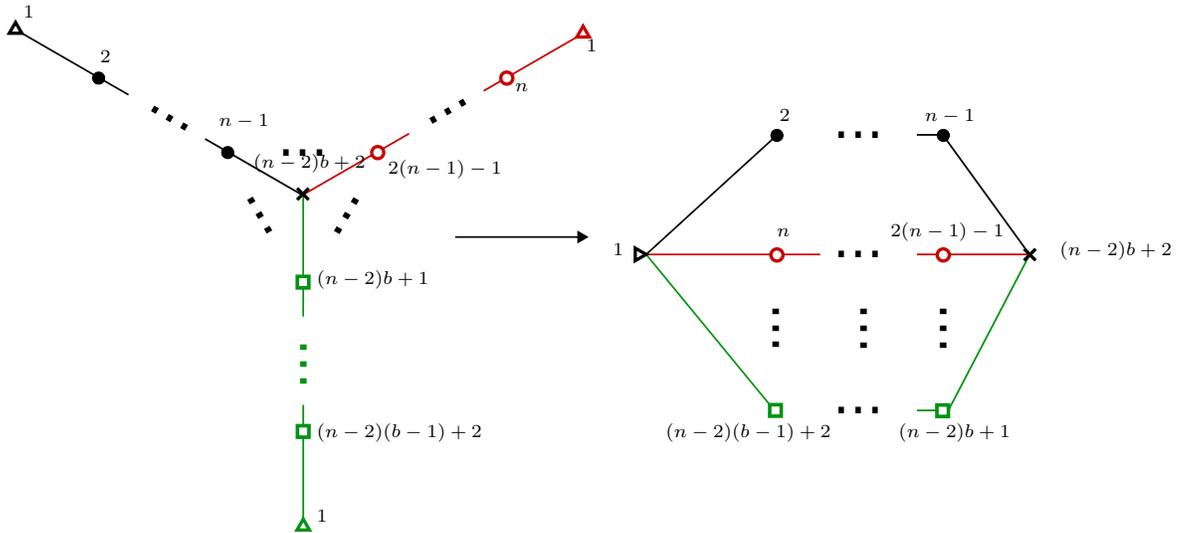
\begin{figure}[h]
		\centering
		\begin{overpic}[width=0.77\textwidth]{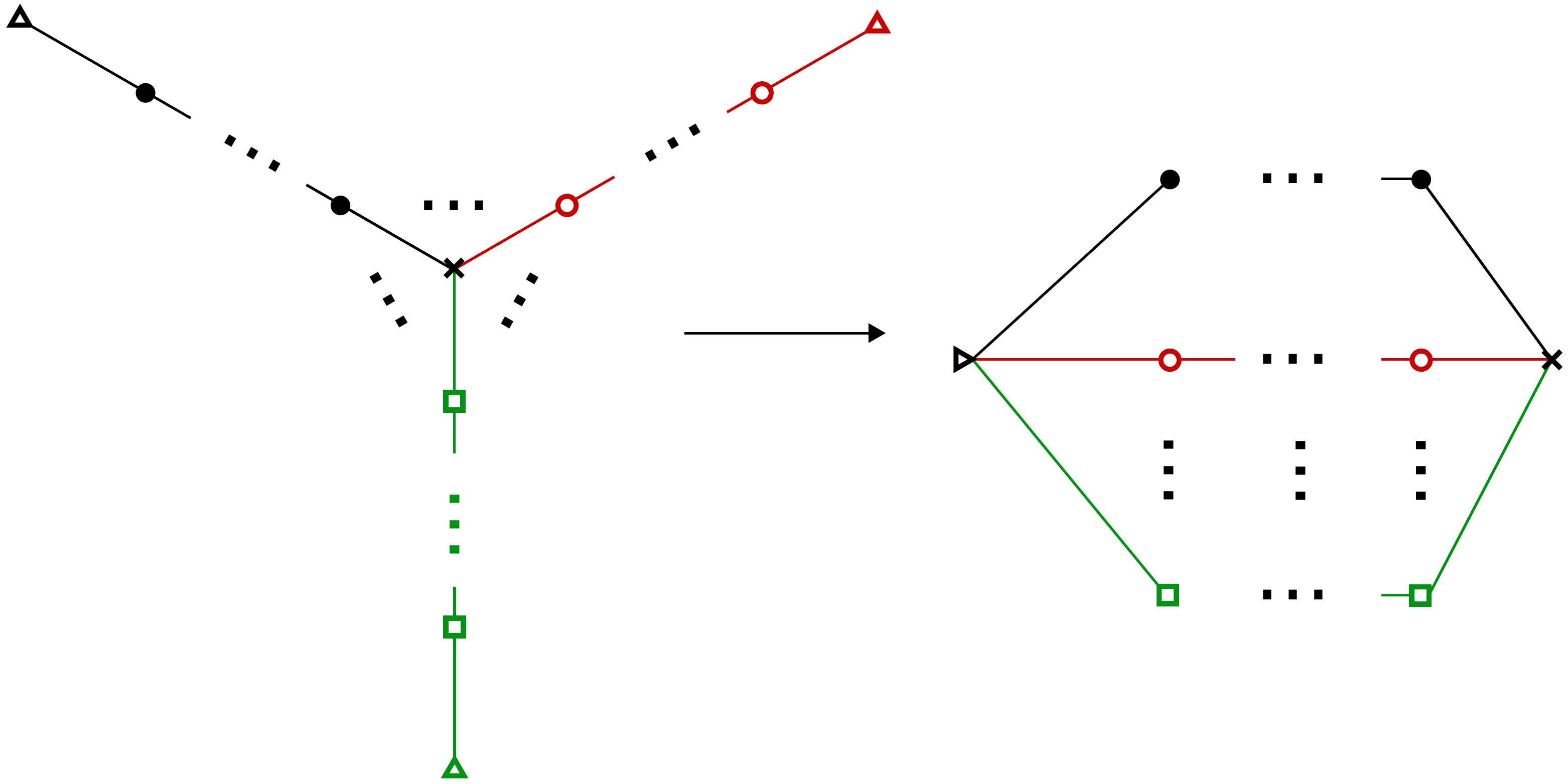}
			\put(231,106){\fontsize{7}{1}$1$}
			\put(294,157){\fontsize{7}{1}$2$}
			\put(349,157){\fontsize{7}{1}$n-1$}
			\put(293,113){\fontsize{7}{1}$n$}
			\put(336,113){\fontsize{7}{1}$2(n-1)-1$}
			\put(251,37){\fontsize{7}{1}$(n-2)(b-1)+2$}
			\put(339,37){\fontsize{7}{1}$(n-2)b+1$}
			\put(400,106){\fontsize{7}{1}$(n-2)b+2$}
			\put(8,196){\fontsize{7}{1}$1$}
			\put(37,179){\fontsize{7}{1}$2$}
			\put(82,155){\fontsize{7}{1}$n-1$}
			\put(221,183){\fontsize{7}{1}$1$}
			\put(194,168){\fontsize{7}{1}$n$}
			\put(146,137){\fontsize{7}{1}$2(n-1)-1$}
			\put(119,5){\fontsize{7}{1}$1$}
			\put(119,37){\fontsize{7}{1}$(n-2)(b-1)+2$}
			\put(119,95){\fontsize{7}{1}$(n-2)b+1$}
			\put(95,139){\fontsize{7}{1}$(n-2)b+2$}
		\end{overpic}
		\caption{Graph formed by merging the outer most vertices of the branches of the graph  described previously.} \label{fig:hain}
	\end{figure}

We will now show that by appending phase terms to graph edges complete suppression of probability amplitudes to vertices of certain graphs can be achieved due to quantum interference similar in nature to that of the Aharonov-Bohm effect \cite{aharonov1959significance}. This extends on work completed by Zimbor\'{a}s et al. where it was shown that complete suppression of information transfer can be achieved to diametrically opposite vertices of even cycles provided the phases satisfy certain conditions \cite{zimboras2013quantum}.  It is noted that breaking TRS is not a necessary condition to observe this phenomenon; however, we will still refer to the quantum walk on graphs with complex edges weightings as a CQW.\\
	\\
	Consider the simple path consisting of $n$ vertices shown in figure \ref{fig:path}. The orthonormal basis associated with this path is $\{ |1\rangle, |2\rangle, ... , |n\rangle \}$ where $|1\rangle$ corresponds to the left most vertex and $|n\rangle$ corresponds to the right most vertex.
	\\
	The corresponding adjacency matrix of this path is given by 
	\begin{align}
	A_{jk}=\begin{cases}
    1 \textrm{ for } |j-k|=1 \\
    0 \textrm{ otherwise}
    \end{cases},
	\end{align}
	where $A$ is an $n \times n$ matrix. We now consider the case where phases have been appended to all edges such that the resulting adjacency matrix is given by
	\begin{align}
	A'_{jk}= \begin{cases}
    e^{-i \alpha_{j-1}} \textrm{ for } j=k+1 \\
    e^{i \alpha_j} \textrm{ for } j=k-1\\ 
    0 \textrm{ otherwise}
    \end{cases}
	\end{align}
	We also introduce the additional phase $\alpha_0=0$ for the sake of completeness. The action of the Hamiltonian is then defined as such for both cases,
	\begin{align}
	H|j\rangle=A|j\rangle = \begin{cases}
	|2\rangle \ \textrm{for} \ j = 1 \\
	|j-1\rangle + |j+1\rangle \ \textrm{for} \ 1<j<n \\
	|n-1\rangle \ \textrm{for} \ j=n \\
	\end{cases},
	\end{align}
	and for the phase appended case
	\begin{align}
	H'|j\rangle=A'|j\rangle = \begin{cases}
	e^{-i\alpha_1}|2\rangle \ \textrm{for} \ j = 1 \\
	e^{i\alpha_{j-1}}|j-1\rangle + e^{-i\alpha_{j}}|j+1\rangle \ \textrm{for} \ 1<j<n \\
	e^{i\alpha_{n-1}}|n-1\rangle \ \textrm{for} \ j=n \\
	\end{cases}.
	\end{align}
	We will let $|\psi(t)\rangle$  and $|\phi(t)\rangle$ denote the state vector of the CTQW and CQW on this path respectively. Considering the case where the walker is initialised at the left-most vertex of the path ($|\psi(0)\rangle = |\phi(0)\rangle=|1\rangle$). We have 
	\begin{align}
	\ket{\psi(t)} & = e^{-iHt} |1\rangle \\
	& = |1\rangle - it|2\rangle - \frac{t^2}{2} \left( \ket{1} + \ket{3} \right)  + \frac{it^3}{6}\left( 2\ket{2} + \ket{4} \right) + \frac{t^4}{24}\left(2\ket{1} + 3\ket{3} + \ket{5} \right) \dots \\
	& = \sum_{j=1}^n \alpha_{j}(t)|j\rangle,
	\end{align}
	and likewise we have
	\begin{align}
	\ket{\phi(t)} & = e^{-iH't} |1\rangle \\
	& = \ket{1} - it e^{-i\alpha_1}\ket{2} - \frac{t^2}{2} \left(\ket{1} + e^{-i(\alpha_1+\alpha_2)} \ket{3} \right) + \frac{it^3}{6}\left( 2e^{-i\alpha_1}\ket{2}  + e^{-i(\alpha_1 + \alpha_2 + \alpha_3)}\ket{4} \right) \\
	&+ \frac{t^4}{24}\left(2\ket{1} + 3 e^{-i(\alpha_1+\alpha_2)}\ket{3} +  e^{-i(\alpha_1+\alpha_2+\alpha_3 + \alpha_4)}\ket{5} \right) \dots \nonumber \\
	& = \sum_{j=1}^n e^{-i\sum_{l=0}^{j-1} \alpha_l} \alpha_{j}(t)|j\rangle \\
	& = \sum_{j=1}^n \beta_j(t)|j\rangle.
	\end{align}
	With a little work it can be shown that the probability amplitudes denoted by $\beta_j(t)$ are related to the probability amplitudes, $\alpha_j(t)$, via the following:
	\begin{align}
	\beta_{j}(t) =  e^{-i\sum_{l=0}^{j-1} \alpha_l} \alpha_{j}(t) \ \textrm{for} \ j=1,2,...,n . \label{eq:cond}
	\end{align}
That is, for a CQW initialised at $|1\rangle$ the probability amplitude at vertex $j$ is found by multiplying the corresponding amplitude for a CTQW by the sum of phases along the edges from vertex $1$ to $j$. \\
	\\
We can generalise  Eq.\eqref{eq:cond} to the case where the initial state for both the CTQW and the CQW is an arbitrary basis state $|k\rangle$. Note that $U_{ij}$ is equivalent to a sum of all paths between $i$ and $j$, and the phases cancel for all paths which are traversed by the walker in both directions. This leaves only the phase components from $i$ to $i+1$, $i+1$ to $i+2$, ... $j-1$ to $j$, which gives us the expression
	\begin{align}
	\beta_{j}(t) =
	\begin{cases}
	e^{-i\sum_{l=k}^{j-1} \alpha_l} \alpha_{j}(t) \ \textrm{for} \ j > k\\
	\alpha_{j}(t) \ \textrm{for} \ j = k\\
	e^{i\sum_{l=j}^{k-1} \alpha_l} \ \textrm{for} \ j < k
	\end{cases}.
	\end{align}
We now consider a graph with $b$ paths each with $n$ vertices joined at their end vertex as shown in figure \ref{fig:type1}. We define a CQW on this graph with an initial state to be superposition of the outermost states of all paths,
\begin{align}
	|\phi(0) \rangle = \frac{\sum_{j=0}^{b-1} | (n-1)j+1  \rangle }{\sqrt{b}}, \label{eq:init}
\end{align}
where graph $G$ is such that $|G|=(n-1)b+1$.  
\\    \\
Let the $p^{th}$ subgraph be the spanning subgraph where all edges are removed apart from the edges that belong to the $p^{th}$ branch . We will let a superscript $p$ denote the state vector of the CQW on the $p^{th}$ subgraph. First we note that the state vector of the entire graph and the $p^{th}$ subgraph for all $p$ must satisfy the Schr\"{o}dinger equation. We have
	\begin{align}
	i\frac{d\ket{\phi(t)}}{dt}=H\ket{\phi(t)} \text{ ~and ~ }  i\frac{d\ket{\phi^p(t)}}{dt}=H^p\ket{\phi^p(t)}.
	\end{align}
Likewise we have for the amplitudes
	\begin{align}
	i\frac{d\beta_{N}(t)}{dt} =\sum_{j=1}^{N}H_{N,j}\beta_j(t) \label{eq:schro}   \text{ ~and ~ }   i\frac{d\beta^p_{N}(t)}{dt}=\sum_{j=1}^{N}H^p_{N,j}\beta^p_{j}(t).
	\end{align}
Given that the vertex set is the same for all subgraphs and the original graph, no two paths have any common edges and that, collectively, the edge sets of all subgraphs form a partition of the edge set of the original graph, we have $H=\sum_{p=1}^b H^p$.\\
	\\
We now consider the initial state $|\phi^p(0)\rangle=|\phi(0)\rangle$ where $|\phi(0)\rangle$ is given in Eq.\eqref{eq:init} above. We can deduce that the vertices of each $p$-subgraph with degree of $0$ (excluding the vertices with initial non-zero probability amplitude) will never be occupied. This implies that
	\begin{align}
	\begin{cases}
	\beta^p_{j}(t) \neq 0 \ \textrm{for} \  (n-1)(p-1) + 2 \leq j \leq (n-1)p   \ \textrm{or} \  j \mod (n-1) = 1 \ \textrm{or} \ j = N. \\
	\beta^p_{j}(t) = 0 \ \textrm{otherwise}.  
	\end{cases}
	\end{align}
	Following this we can deduce that 
	\begin{align}
	H^k_{N,j}\beta^p_{j}(t) = 0 \ \textrm{for} \ k \neq p \ \textrm{and for all} \ j.
	\end{align} 
	The reason for this is because the only non-zero entry of $H^k_{N,j}$ is the $(n-1)k^{th}$ entry but $\beta^p_{(n-1)k}(t)=0$.\\
	\\
	We now have
	\begin{align}
	\sum_{p=1}^b i\frac{d\beta^p_{N}(t)}{dt} &=\sum_{p=1}^b \sum_{j=1}^{N}H^p_{N,j}\beta^p_j(t), \end{align}
and hence
\begin{align}
	i\frac{d \sum_{p=1}^b \beta^p_{N}(t)}{dt} &=\sum_{p=1}^b\sum_{j=1}^{N}  \left(H^p_{N,j}\beta^p_j(t) +\sum_{k \neq p} H^k_{N,j}\beta^p_j(t) \right)=\sum_{j=1}^{N} H_{N,j} \sum_{p=1}^b \beta^p_j(t).
	\end{align}
	Comparing this with Eq.\eqref{eq:schro} and since $\ket{\phi(0)} = \ket{\phi^p(0)} $ we have
	\begin{align}
	\beta_{N}(t)=\sum_{p=1}^b \beta^p_{N}(t)
	\end{align}
	up to a differing normalization constant. This means that 
	\begin{align}
	\beta^p_{N}(t) = e^{-i\sum_{l=0}^{n-1} \alpha^p_l} \alpha^p_{N}(t).
	\end{align}
	Since $ \alpha^p_{N}(t)$ is same for all $p$-subgraphs we will drop the superscript $p$ for this term. Note that this is not the same as the probability amplitude of the original graph. We then have
	\begin{align}
	\beta_{N}(t) &=\sum_{p=1}^b \beta^p_{N}(t), 
\end{align}
and hence
\begin{align}    
    \beta_{N}(t)	& = \alpha_{N}(t)\sum_{p=1}^b  e^{-i\sum_{l=0}^{n-1} \alpha^p_l} .
	\end{align}
	If we wish to suppress information transfer to vertex $|N \rangle$ we must have $\beta_{N}(t)	=0$ for all $t$. Therefore we must have
	\begin{align}
	\sum_{p=1}^b  e^{-i\sum_{l=0}^{n-1} \alpha^p_l} = 0
	.  \label{eq:zerotran} \end{align}
Hence, given a graph which consists of $b$ paths all with $n$ vertices and all paths joined by their end vertex, a CQW is initialised according to Eq.\eqref{eq:init} can achieve zero state transfer to the merged end vertex provided the above condition given in Eq.\eqref{eq:zerotran} is satisfied. \\
\\
We can further extend the zero transfer phenomenon to graphs obtained by merging the outer most vertex of each path as shown in figure \ref{fig:hain}. Let the initial state of the CQW on this graph be $|\phi(0)\rangle= |1\rangle$, one can see that, given the relationship between the two types of graph, the initial states are equivalent such that zero transfer will occur to $|(n-2)b+2\rangle$ (labeled as $|(n-1)b+1\rangle$ in the former case).\\
\\
\section{Examples}
\begin{figure}[h] \centering
		\begin{overpic}[width=0.95\textwidth]{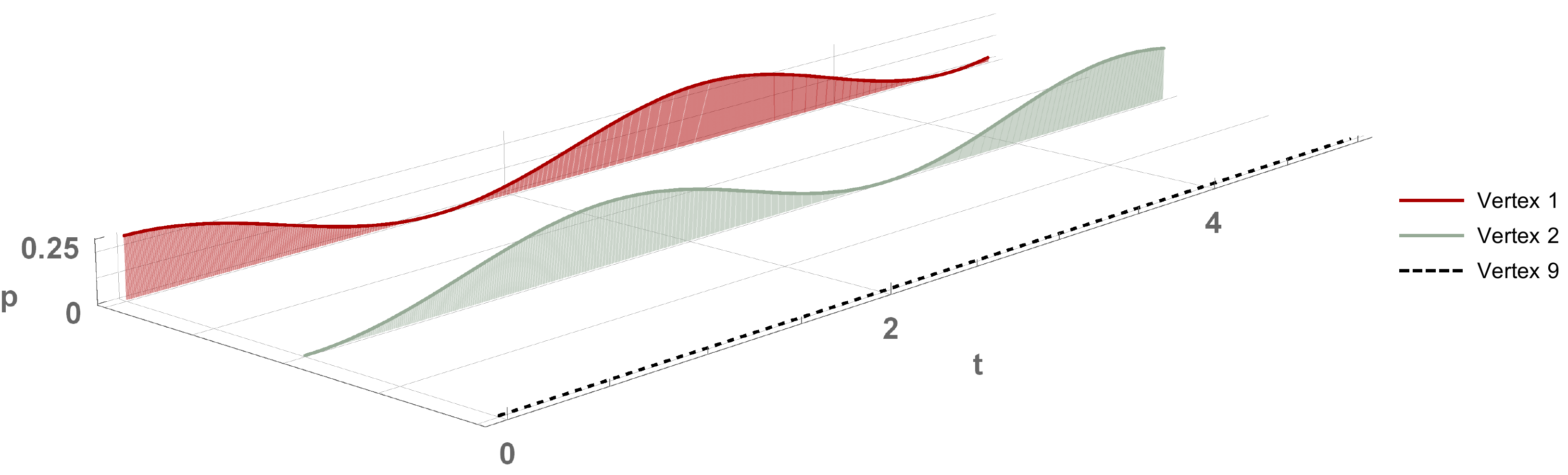}
		\end{overpic}
		\caption{Probability of observing the walker at a respective vertex during times, $t=0$ to $t=5$, for graph of first type with parameters $b=4$ and $n=3$.} \label{fig:restype1}
	\end{figure}

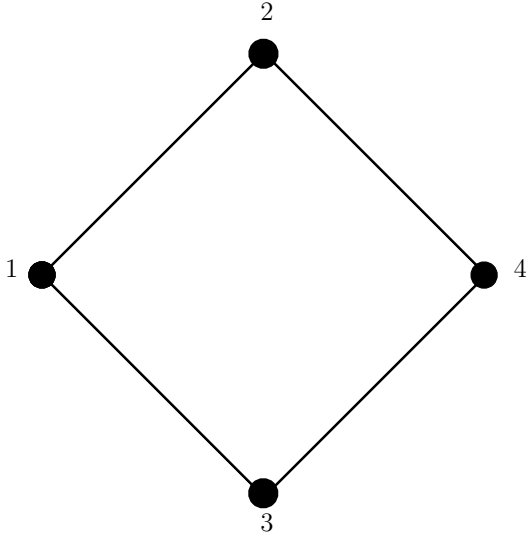
\begin{figure}[h] \centering
\scalebox{0.7}{
		\begin{overpic}[width=0.6\textwidth]{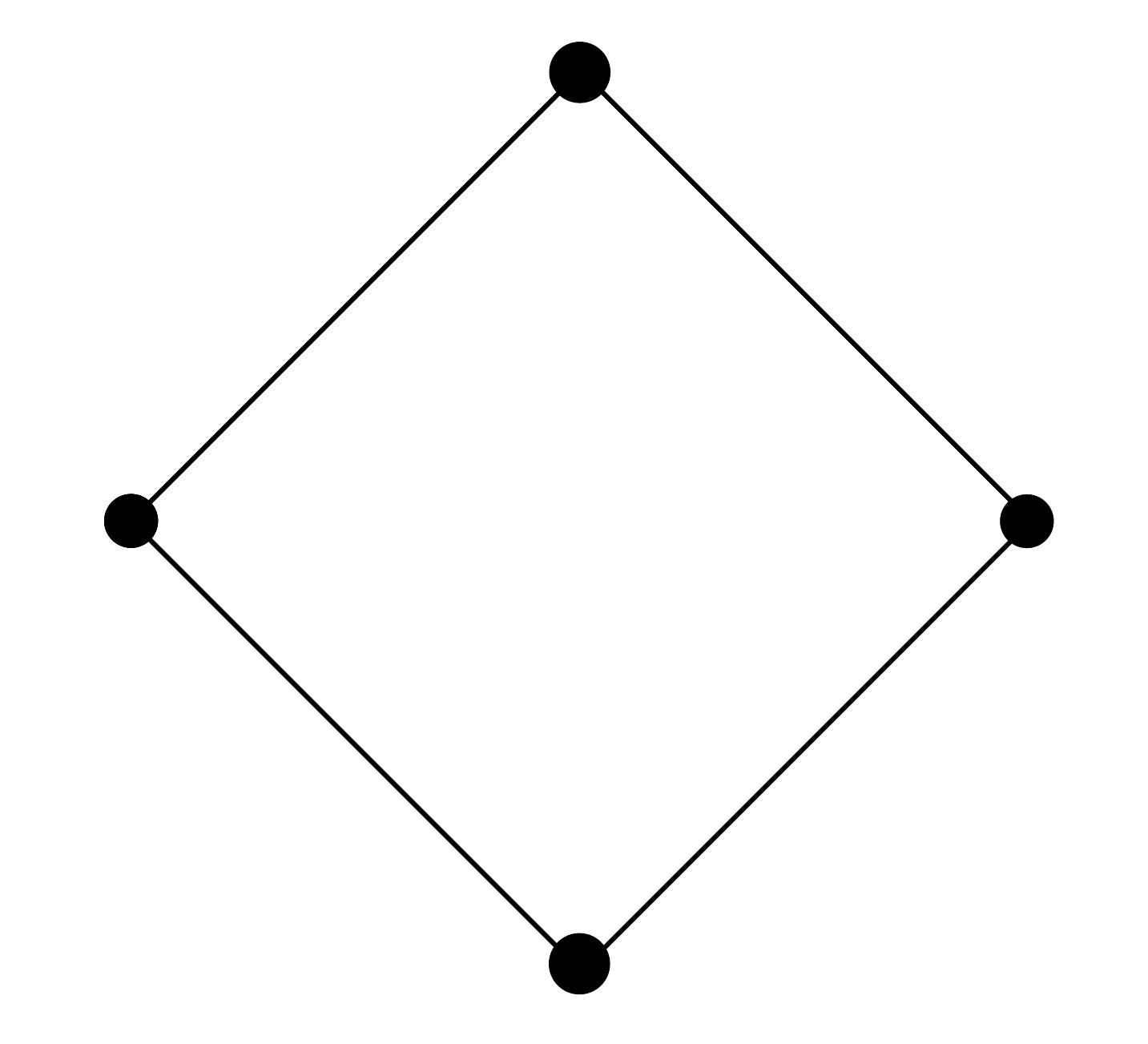}
			\put(15, 138){\fontsize{14}{1}$1$}
			\put(153,277){\fontsize{14}{1}$2$}
			\put(153, 1){\fontsize{14}{1}$3$}
			\put(290,138){\fontsize{14}{1}$4$}
		\end{overpic}
        }
		\caption{Cycle with 4 vertices, $C_4$.} \label{fig:C4}
	\end{figure}

\begin{figure}[h] \centering
	\begin{overpic}[width=0.95\textwidth]{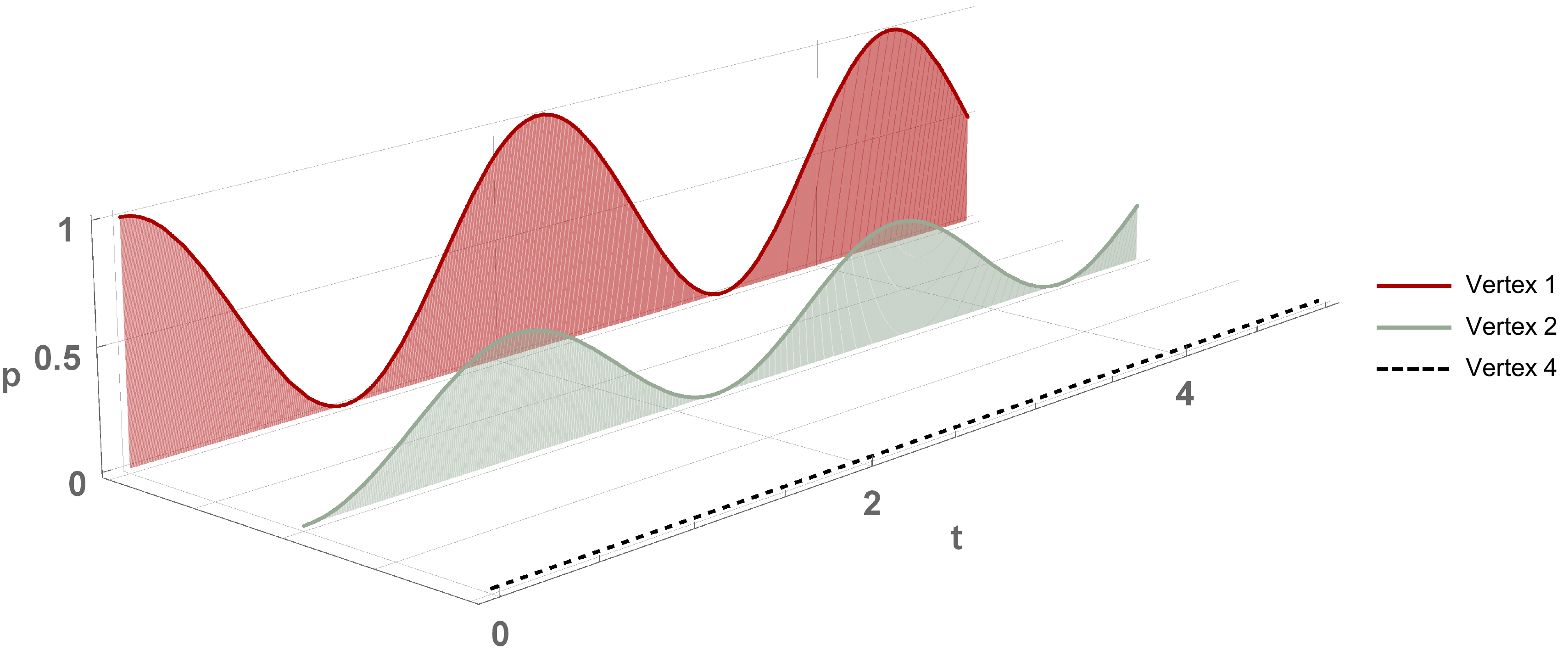}
	\end{overpic}
	\caption{Probability of observing the walker at a respective vertex during times, $t=0$ to $t=10$, for $C_4$.} \label{fig:resC4}
\end{figure}

\begin{figure}[h]  \centering
		\begin{overpic}[width=0.60\textwidth]{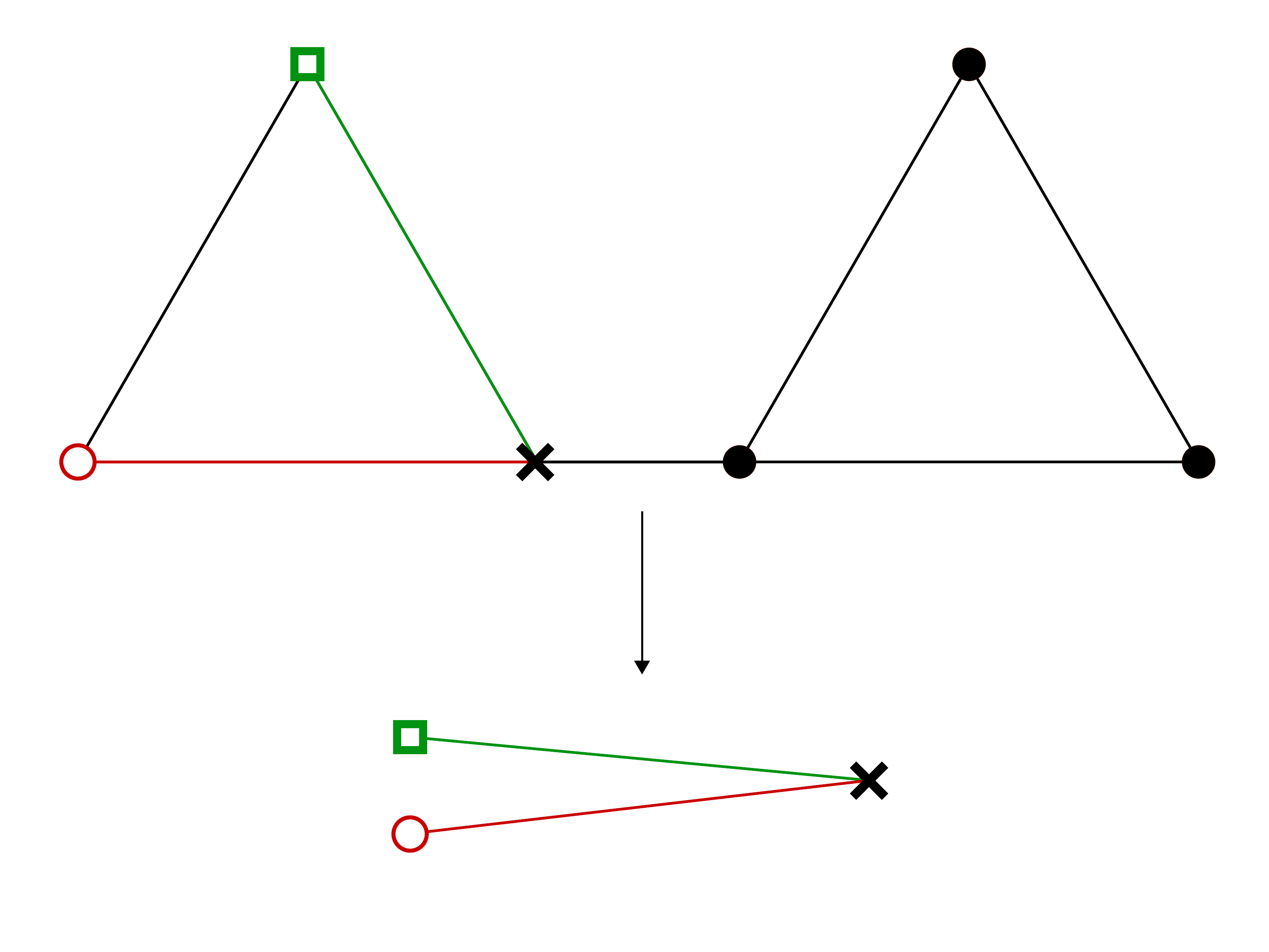}
			\put(-1, 124){$1$}
			\put(70, 224){$2$}
			\put(136, 124){$3$}
			\put(166, 124){$4$}
			\put(231, 224){$5$}
			\put(299, 124){$6$}
			\put(82, 49){$2$}
			\put(82, 25){$1$}
			\put(217, 39){$3$}
		\end{overpic}
		\caption{Miscellaneous graph and corresponding path decomposition.} \label{fig:misc}
	\end{figure}

\begin{figure} \centering
		\begin{overpic}[width=0.95\textwidth]{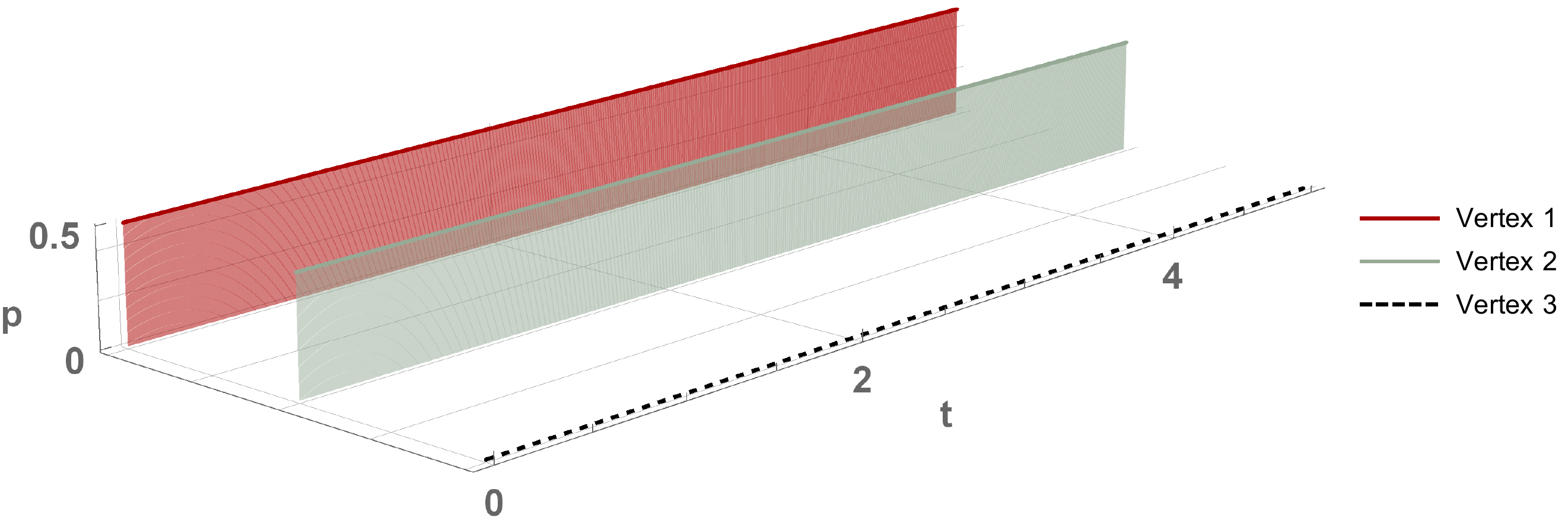}
		\end{overpic}
		\caption{Probability of observing the walker at a respective vertex during times, $t=0$ to $t=10$, for miscellaneous graph.} \label{fig:resmisc}
	\end{figure}

	Consider the graph shown in figure \ref{fig:type1ex} which is of the first type of graph described above with parameters $b=4$ and $n=3$. The adjacency matrix of this graph is 
	\begin{align}
	A = \begin{bmatrix}
	0 & 1 & 0 & 0 & 0 & 0 & 0 & 0 & 0 \\
	1 & 0 & 0 & 0 & 0 & 0 & 0 & 0 & 1 \\
	0 & 0 & 0 & 1 & 0 & 0 & 0 & 0 & 0 \\
	0 & 0 & 1 & 0 & 0 & 0 & 0 & 0 & 1 \\
	0 & 0 & 0 & 0 & 0 & 1 & 0 & 0 & 0 \\
	0 & 0 & 0 & 0 & 1 & 0 & 0 & 0 & 1 \\
	0 & 0 & 0 & 0 & 0 & 0 & 0 & 1 & 0 \\
	0 & 0 & 0 & 0 & 0 & 0 & 1 & 0 & 1 \\
	0 & 1 & 0 & 1 & 0 & 1 & 0 & 1 & 0\\
	\end{bmatrix}.
	\end{align}
    	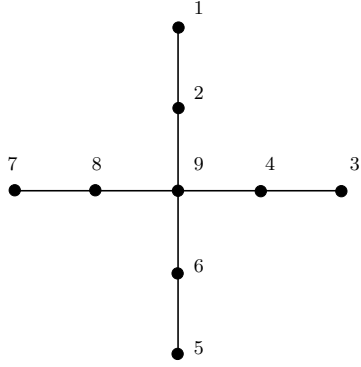
\begin{figure}[htb] \centering
    \scalebox{0.82}{
		\begin{overpic}[width=0.33\textwidth]{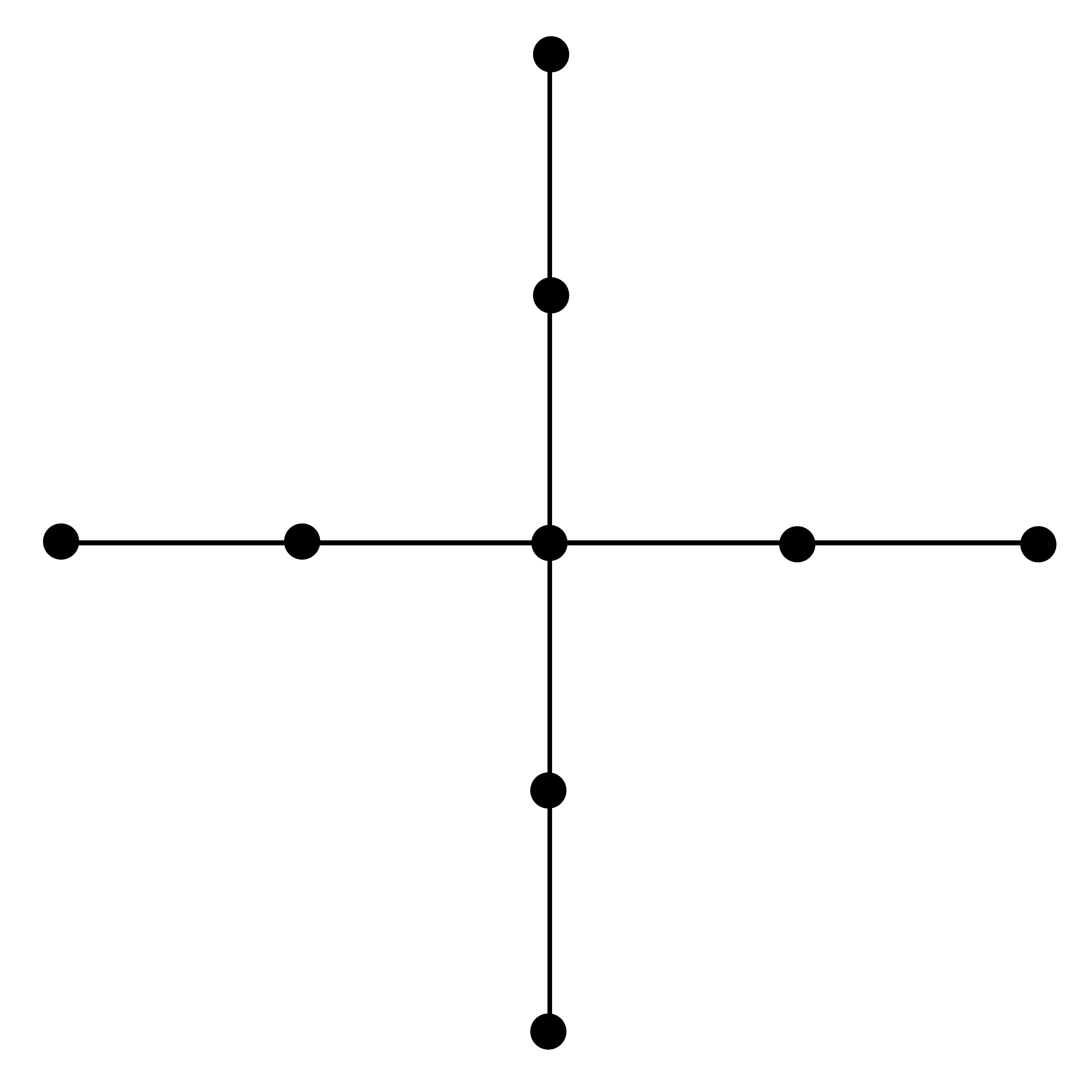}
			\put(92,166){\fontsize{9}{1}$1$}
			\put(92,127){\fontsize{9}{1}$2$}
			\put(164,94){\fontsize{9}{1}$3$}
			\put(125,94){\fontsize{9}{1}$4$}
			\put(92,9){\fontsize{9}{1}$5$}
			\put(92,47){\fontsize{9}{1}$6$}
			\put(6,94){\fontsize{9}{1}$7$}
			\put(45,94){\fontsize{9}{1}$8$}
			\put(92,94){\fontsize{9}{1}$9$}			
		\end{overpic}
}
		\caption{First type of graph with parameters $b=4$ and $n=3$.} \label{fig:type1ex}
	\end{figure}\\
	\\
	We will choose to append the terms $e^{i  \pi /2 }$ to the directed edge $(1,2)$, $e^{i  \pi }$ to the directed edge $(3,4)$, $e^{i3\pi/2}$ to the directed edge $(5,6)$ and their respective complex conjugates in the opposing directions. The adjacency matrix then becomes 
	\begin{align}
	A' = \begin{bmatrix}
	0 & i & 0 & 0 & 0 & 0 & 0 & 0 & 0 \\
	-i & 0 & 0 & 0 & 0 & 0 & 0 & 0 & 1 \\
	0 & 0 & 0 & -1 & 0 & 0 & 0 & 0 & 0 \\
	0 & 0 & -1 & 0 & 0 & 0 & 0 & 0 & 1 \\
	0 & 0 & 0 & 0 & 0 & -i & 0 & 0 & 0 \\
	0 & 0 & 0 & 0 & i & 0 & 0 & 0 & 1 \\
	0 & 0 & 0 & 0 & 0 & 0 & 0 & 1 & 0 \\
	0 & 0 & 0 & 0 & 0 & 0 & 1 & 0 & 1 \\
	0 & 1 & 0 & 1 & 0 & 1 & 0 & 1 & 0\\
	\end{bmatrix}.
	\end{align}
The initial state of the CQW is chosen to be 
	\begin{align}
	|\phi(0) \rangle = \frac{|1\rangle + |3\rangle + |5\rangle + |7\rangle}{2}.
	\end{align}    
Following this, one can check to see that the initial state and zero transfer condition stated in Eq.\eqref{eq:init} and Eq.\eqref{eq:zerotran} respectively are satisfied. Figure \ref{fig:restype1} shows the probability of observing the walker at vertices 1,2 and 9 for various times where other vertices have been excluded due to symmetry of the walk.  As it can be seen the probability of observing the walker at $|9\rangle$ is 0 for all $t$ shown.
\\
Next consider the cycle graph, $C_4$, shown in figure \ref{fig:C4}. $C_4$ falls under the category of the graph of the second type described previously with parameters $b=3$ and $n=2$. We let the initial state of the CQW on this graph be $|\phi(0)\rangle  = |1\rangle$. We now append $e^{i \pi}$ to the directed edge $(1,2)$ and its complex conjugate in the opposing direction. It can be shown that these phases satisfy the conditions given in Eq.$\eqref{eq:zerotran}$ for zero transfer to vertex $|4\rangle$. \\
\\
\\
\\
The corresponding phase appended adjacency matrix of this CQW is given by 
\begin{align}
	A' = \begin{bmatrix} 0 & e^{i \pi} & 1 & 0 \\
	e^{-i \pi} & 0 & 0 & 1 \\
	1 & 0 & 0 & 1 \\
	0 & 1 & 1 & 0 \end{bmatrix}.
\end{align}
With some working one can show that for the time evolution operator corresponding to this adjacency matrix the TRS condition is still satisfied, and is in fact satisfied for all bipartite graphs (which includes all cycle graphs of even length) regardless of appended phase terms \cite{lu2016chiral}. However, by using similar methods to the example of $C_4$ (which itself is an even cycle graph), all even cycle graphs can achieve zero transfer to any of its vertices. At first this may seem counter-intuitive since we are achieving suppression of information transfer whilst TRS still holds. However, it is noted that TRS only implies that the STP between two distinct vertices is the same in either direction. In the case where zero transfer occurs there is an STP of 0 along one direction of the edge connected to this vertex and an STP of 0 in the other direction of this same edge, hence the TRS condition is not violated. Figure \ref{fig:resC4} shows the probability of observing the walker at vertices 1,2 and 4 for various times where vertex 3 has been excluded to due symmetry of the walk. As can clearly be seen, the probability amplitude of $|4\rangle$ has been completely suppressed.
\\
\\
Finally we show that some edges of graphs can play a passive role in terms of exhibiting this phenomenon. Consider a graph with the following adjacency matrix 
\begin{align}
	A=\begin{bmatrix}
	0 & 1 & 1 & 0 & 0 & 0\\
	1 & 0 & 1 & 0 & 0 & 0\\
	1 & 1 & 0 & 1 & 0 & 0\\
	0 & 0 & 1 & 0 & 1 & 1\\
	0 & 0 & 0 & 1 & 0 & 1\\
	0 & 0 & 0 & 1 & 1 & 0 
	\end{bmatrix},
\end{align}
and let the initial state of this CQW be 
\begin{align}
	|\phi(0)\rangle  = \frac{|1\rangle + |2\rangle}{\sqrt{2}}.
\end{align}
If we wanted to achieve zero transmission to $|3\rangle$ (and subsequently $|4\rangle$ and higher) we think of $1\sim 3$ and $2\sim 3$ as paths of 2 vertices each merged at vertex $3$ as shown in figure \ref{fig:misc}. Note here $1\sim 2$ does not play any active role due to the symmetry of the graph.
By appending $e^{i\pi}$ to directed edge $(1,3)$ and its complex conjugate in the opposing direction the adjacency matrix becomes
	\begin{align}
	A'=\begin{bmatrix}
	0 & 1 & -1 & 0 & 0 & 0\\
	1 & 0 & 1 & 0 & 0 & 0\\
	-1 & 1 & 0 & 1 & 0 & 0\\
	0 & 0 & 1 & 0 & 1 & 1\\
	0 & 0 & 0 & 1 & 0 & 1\\
	0 & 0 & 0 & 1 & 1 & 0 
	\end{bmatrix}.
	\end{align}
Figure \ref{fig:resmisc} shows the probability of observing the walker at vertices 1,2 and 3 for various times.  It can clearly be seen that the probability amplitude of $|3\rangle$ has been completely suppressed and although excluded from the figure, one can check to see that zero transmission is achieved to vertices 4,5 and 6 as well. In addition to this we observe the fact that the system remains in its initial state for the entirety of the CQW.  	
\\
\\
\\
\\
\\
\\
\\
\\
\\
\\
\\
\\
\\
\\
\\
\\
\\
\\
\\
\section{Addition of Decoherence and Application}
\begin{figure}[h] \centering
		\begin{overpic}[width=1\textwidth]{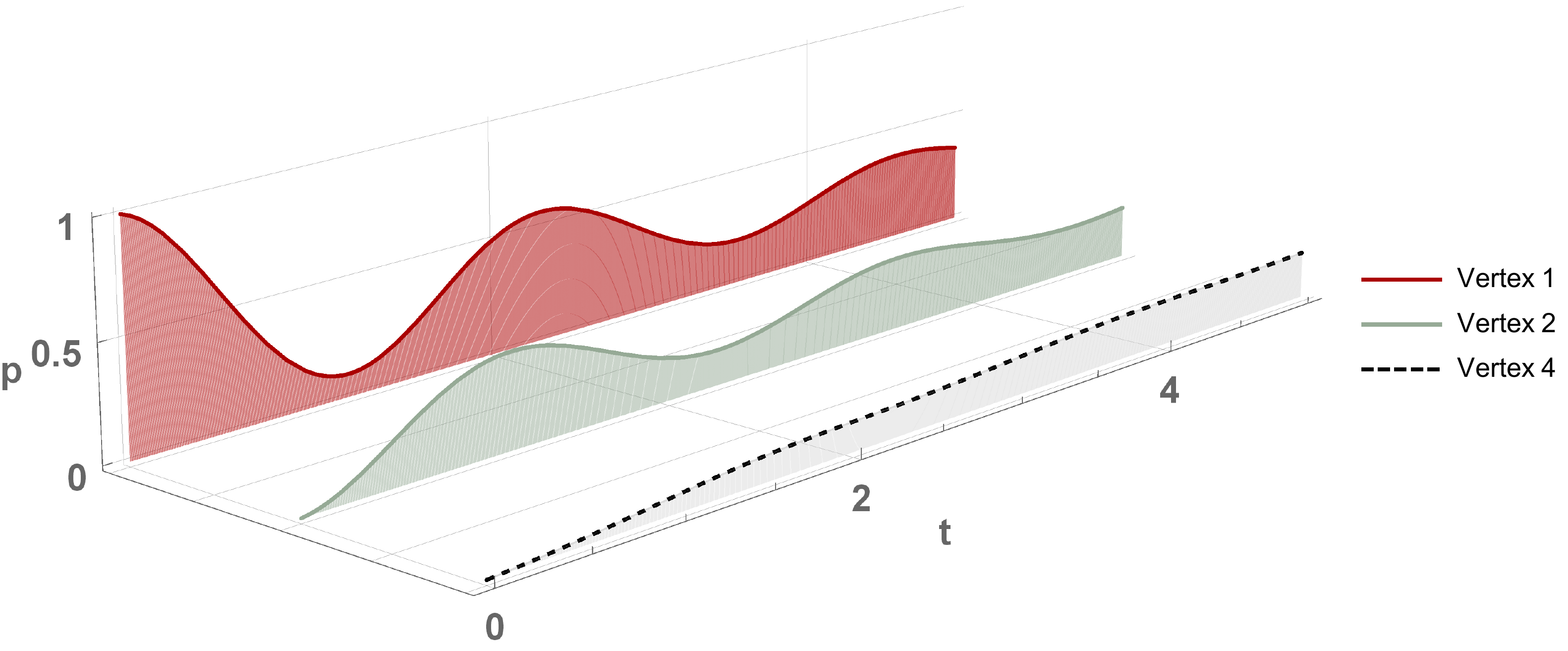}
		\end{overpic}
		\caption{Probability of observing the walker at a respective vertex during times, $t=0$ to $t=10$, for $C_4$ with scattering, dissipation and dephasing decoherence.} \label{fig:scatres}
	\end{figure}

\begin{figure}[h] \centering
	\begin{overpic}[width=0.28\textwidth]{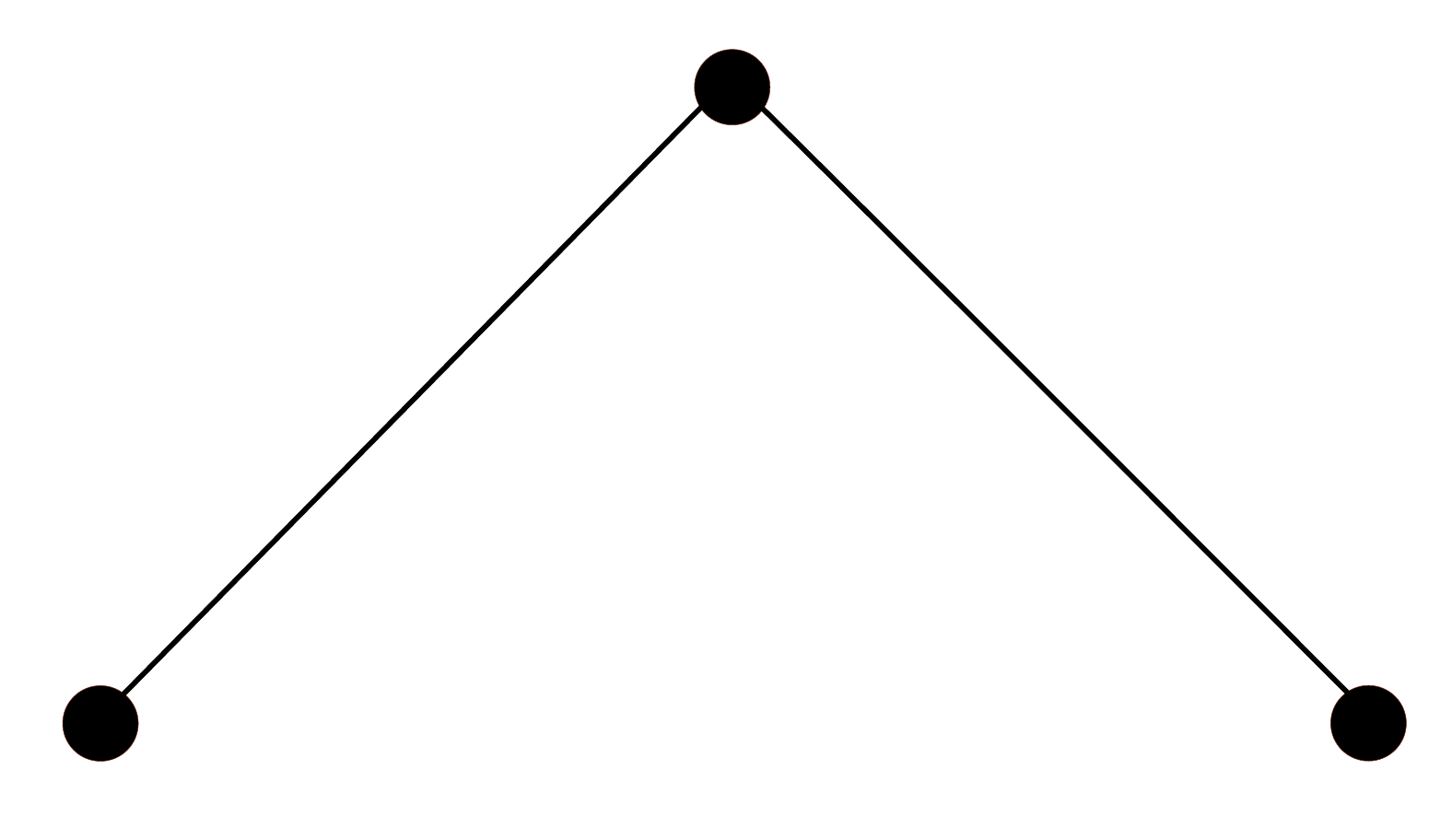}
		\put(7, -1){$1$}
		\put(137, -1){$3$}
		\put(71, 87){$2$}	
	\end{overpic}
	\caption{A graph with three vertices.} \label{fig:measure}
\end{figure}

To assess its robustness to decoherence, the phenomenon was modeled in an open system using quantum stochastic walks (QSWs). The QSW is a generalisation of the CTQW using the Lindblad equation \cite{whitfield2010quantum} which can be used to incorporate directionality in a quantum walk \cite{sanchez2012quantum}. The Lindblad equation is given by
	\begin{align}
	\frac{d\hat{\rho}}{dt}=-(1-\omega)i[\hat{H},\hat{\rho}(t)]+\omega \sum_{k=1}^K \left( \hat{L}_k \hat{\rho}(t)\hat{L}^\dag_k-\frac{1}{2}\left(\hat{L}^\dag_k \hat{L}_k \hat{\rho}(t) + \hat{\rho}(t) \hat{L}_k^\dag \hat{L}_k \right) \right).
	\end{align}
Here the classical part of the equation is used to describe the non-unitary part of the stochastic walk which is used to incorporate directionality. To simulate environmental interactions decoherence was introduced to the system in the form of scattering, dissipation and dephasing through the use of the Lindblad operators, $L_k$, in QSWs. This was done using the `QSWalk' package by Falloon et al. \cite{falloon2017qswalk}, which computes quantum stochastic walks on graphs using the Mathematica system.\\
\\
Figure \ref{fig:scatres} shows the probability distribution once again for $C_4$ with the same complex edge weights given in the example before with decoherence introduced to the system. It can be seen that when decoherence is introduced to the system the zero transfer phenomenon no longer occurs. In fact as $t$ grows large the quantum walk eventually becomes classical in its behaviour, on a time scale determined by $\omega$. \\
\\
Hence, this phenomenon does not occur when decoherence is taken into consideration. Given that no physical system can exist in perfect isolation this then eliminates many possibilities of utilising this phenomenon for information processing/transfer applications. However it is proposed that this phenomenon can be used to detect the presence of decoherence in a quantum system.\\
\\
Consider the graph shown in figure \ref{fig:measure}. 
We let the initial state of a CQW on this graph be 
	\begin{align}
		|\phi(0)\rangle = \frac{|1\rangle + |3\rangle}{\sqrt{2}},
	\end{align}
and will append the phase term $e^{i\pi}$ to the directed edge $(1, 2)$ and its complex conjugate in the opposing direction. The resulting adjacency matrix will then be given by 
\begin{align}
	A' = \begin{bmatrix}
	0 & -1 & 0 \\
	-1 & 0 & 1 \\
	0 & 1 & 0  
	\end{bmatrix}.
\end{align}
One can see that the following initial conditions and edge weights are such that zero transfer will be achieved at $|2\rangle$. Hence if the walker is measured to be at $|2\rangle$ and little time has elapsed for the CQW this indicates that there is likely a large magnitude of decoherence present.\\
\\
\\
Classically, for a graph with $n$ vertices, when decoherence is introduced to the system the probability of measuring the walk at any vertex approaches an equal superposition, i.e.\ as $t \rightarrow \infty$ the probability of measuring the walker at any vertex is $1/n$. \\
\begin{figure}[h] \centering
	\begin{overpic}[width=0.74\textwidth]{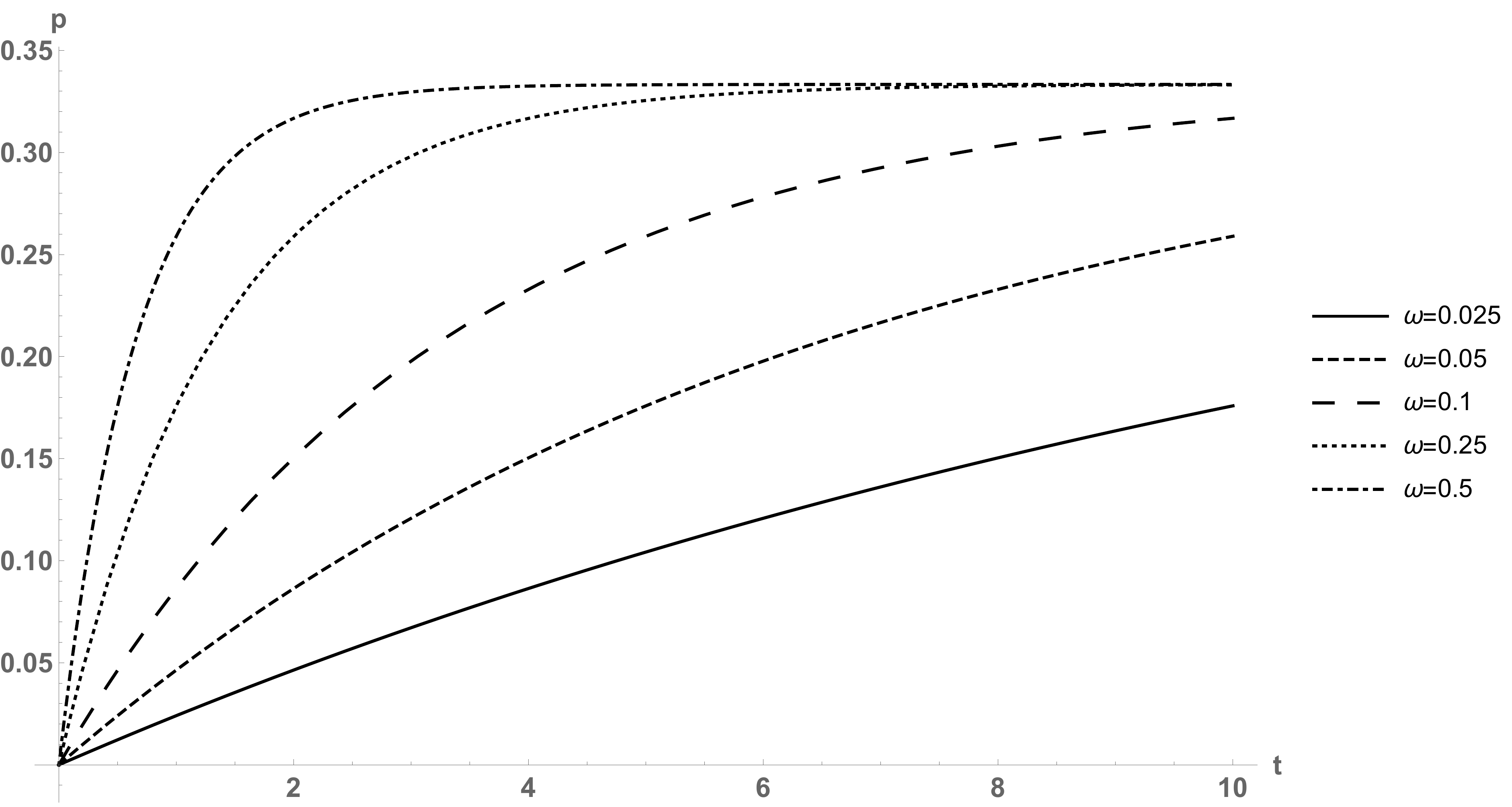}
	\end{overpic}
	\caption{Probability of measuring the walker at vertex 2 for various values of $\omega$.} \label{fig:measureres}
\end{figure}\\
\\
Figure \ref{fig:measureres} gives the probability of observing the walker at vertex 2 over time for various values of $\omega$. Using this data as reference one can then obtain an estimate of the magnitude of deocherence in a quantum system using the following. First one obtains a sample of measurements of the above quantum walk at a given fixed time to obtain the probability of observing the walker at vertex 2 at a given time. Next, one uses the above table to find a corresponding value of $\omega$ and hence an estimate of the magnitude of decoherence in the system.
\section{Conclusion and further work}
Continuous time quantum walks have been extensively studied in the domain of unweighted, undirected graphs. In this paper we show that, by using chiral quantum walks, many graphs can exhibit a zero transfer phenomenon if they can be treated as a union of paths such that all paths collectively satisfy a zero phase condition.\\
\\
The conditions presented are general in the sense that we can observe the zero transfer phenomenon in a variety of graphs. Note that it is not necessary to break the time reversal symmetry condition in order for zero transfer to be observed. Hence an interesting topic of further research is the exploration of phenomena or applications that can be formulated by using complex-valued quantum walk Hamiltonians without the necessity to break the time reversal symmetry condition. \\
\\
Finally, we show that this phenomenon is not robust to environmental interactions by introducing decoherence in the form of scattering, dissipation and dephasing via a quantum stochastic walk. We utilise this property to construct a simple method for estimating the magnitude of decoherence in a given quantum system. We conclude by asking if there are other applications or protocols that will benefit by utilising the zero transfer phenomenon.

\section*{Acknowledgements}
We would like to thank Aeysha Khalique for proof reading the manuscript and providing constructive suggestions.


%

\end{document}